\documentclass[useAMS,aas_macros]{mn2e}
\usepackage{aas_macros}
\usepackage{amssymb}
 \usepackage{epsfig}
 \usepackage{aastex_hack}
 \usepackage{multirow}
 \usepackage{longtable}

\newcommand{\etal}{et~al.~}
\newcommand{\Sersic}{S\'{e}rsic }

\newcommand{\kms}{\ifmmode\,{\rm km}\,{\rm s}^{-1}\else km$\,$s$^{-1}$\fi}
\newcommand{\magarc}{\ifmmode {{{{\rm mag}~{\rm arcsec}}^{-2}}}
             \else {{{mag}$~${arcsec}$^{-2}$}}
             \fi}
                                                                                
\def\arcsec {\ifmmode ^{''}\else $^{''}$\fi} 

\def\Eq#1{Eq.~(\ref{eq:#1})}

\def\se#1{\S\ref{sec:#1}}
\def\Fig#1{Fig.~\ref{fig:#1}}

\def\be{\begin{equation}}
\def\ee{\end{equation}}
\def\ifm#1{\relax\ifmmode#1\else$\mathsurround=0pt #1$\fi}

\def \spose#1{\hbox to 0pt{#1\hss}}
\def \lta{\mathrel{\spose{\lower 3pt\hbox{$\sim$}}
     \raise 2.0pt\hbox{$<$}}}
\def \gta{\mathrel{\spose{\lower 3pt\hbox{$\sim$}}
     \raise 2.0pt\hbox{$>$}}}
\def\ltsima{$\; \buildrel < \frac \sim \;$}
\def\lsim{\lower.5ex\hbox{\ltsima}}
\def\gtsima{$\; \buildrel > \frac \sim \;$}
\def\gsim{\lower.5ex\hbox{\gtsima}}

\def\kms{\ifmmode\,{\rm km}\,{\rm s}^{-1}\else km$\,$s$^{-1}$\fi}

\def\re  {r_{e}}

\def\muc {\mu_0}

\def \V22{V_{2.2}}

\def \ion#1#2{#1{\footnotesize{#2}}\relax}

\def \hi {\ion{H}{I} }

\title{Bulge-Disk Decompositions and Structural Bimodality 
of Ursa Major Cluster Spiral Galaxies}

\author[M. McDonald, S. Courteau \& R.~B. Tully]{Michael McDonald$^{1,*}$, 
 St\'{e}phane Courteau$^1$, \& R. Brent Tully$^2$\\
\\
$^1$Department of Physics, Engineering Physics and Astronomy, 
 Queen's University, Kingston, ON, Canada\\
$^*$Currently at University of Maryland, College Park, MD\\
$^2$Institute for Astronomy, University of Hawaii, 2680 Woodlawn Drive, Honolulu, HI\\
mcdonald@astro.umd.edu, courteau@astro.queensu.ca, tully@ifa.hawaii.edu}

\begin{document}

\pagerange{\pageref{firstpage}--\pageref{lastpage}} \pubyear{2009}

\maketitle 
 
\label{firstpage}

\begin{abstract}
We present bulge and disk (B/D) decompositions of existing $K^\prime$ 
surface brightness profiles for 65 Ursa Major cluster spiral galaxies.  
This improves upon the disk-only fits of Tully et al. (1996).
The 1996 disk fits were used by Tully \& Verheijen (1997) for their 
discovery of the bimodality of structural parameters in the UMa 
cluster galaxies.  It is shown that our new 1D B/D decompositions 
yield disk structural parameters that differ only slightly 
from the basic fits of Tully et al. and evidence for structural bimodality 
of UMa galaxies is maintained.  Our B/D software for the decomposition 
of 1D surface brightness profiles of galaxies uses a non-linear 
minimization scheme to recover the best fitting \Sersic bulge 
and exponential disk while accounting for the possible presence 
of a compact nucleus and spiral arms and for the effects of seeing
and disk truncations.  In agreement with Tully \& Verheijen,
we find that the distribution of near-infrared disk central surface 
brightnesses is bimodal with an F-test confidence of 80\%. There is also strong evidence for a local minimum in the luminosity function
at $M_{K^\prime} \simeq -22$. A connection between the brightness bimodality and a dynamical bimodality, based on new \ion{H}{I} line widths, is identified.
The B/D parameters are presented in an Appendix.
\end{abstract}

                                                                                
\section{Introduction}\label{sec:intro}

The discovery of structural bimodality of Ursa Major (hereafter UMa)
cluster galaxies by Tully \& Verheijen (1997; hereafter TV97) remains
of tremendous interest in light of its implications for galaxy formation 
models and the fact that it still lacks, a full decade later, a firm 
theoretical explanation. Tully et al. (1996; hereafter T96) reported 
deep multiband $BRI$ and $K^\prime$ imaging of a complete sample of 
65 UMa cluster galaxies.  The analysis of these data by TV97 revealed
a striking bimodality in the distribution of inclination-corrected 
disk central surface brightnesses, $\muc^i$.  Rather than showing 
a monotonic range of values starting at the Freeman 
peak\footnote{Freeman (1970) found a very narrow peak in the 
distribution of disk central surface brightnesses at $\mu_0^B=21.65$
for a sample of 28 bright spiral and lenticular galaxies.} for 
brightest galaxies, the distribution of $\muc^i$ 
found by TV97 showed two distinct peaks describing low surface brightness
(hereafter LSB) galaxies and high surface brightness (hereafter HSB)
galaxies respectively.  Using the deep IR imaging alone, which is 
largely insensitive to dust extinction, TV97 found that UMa 
spiral galaxies seem to avoid disk central surface brightnesses 
around $\mu_{0,K} \simeq 18.5$ \magarc. 

T96 fitted exponential disks to the light profiles of the 65 UMa 
galaxies.  These fits yield a value for $\muc$ and the disk scale 
length, $h$.  The inclination correction for $\muc$ used by TV97 
included both geometric and extinction terms, as follows:
\be
\muc^i=\muc-2.5C^\lambda\log(b/a)
\label{eq:inc_corr}
\ee 
where $C^\lambda$ is the wavelength-dependent extinction correction
term and $b/a$ is the axial ratio.  For NIR imaging, and LSB galaxies 
at all wavelengths, it was assumed that $C^\lambda=1$; other values 
for $C^\lambda$ at other dust-sensitive wavelengths for HSB galaxies 
were used by TV97.  
At $K^{\prime}$, the distribution of inclination-corrected  
$\muc^i$ shows two distinct peaks at 17.3 $K^{\prime}$ \magarc 
for HSB galaxies and 19.7 $K^{\prime}$ \magarc for LSB galaxies.  
The bimodality of the $\muc$ distribution was not convincingly observed at 
optical (BRI) bands but correcting the optical surface brightness 
profiles for inclination and extinction effects reveals a 
surface brightness bimodality in these bands as well.  TV97 
stressed it could not be excluded that the SB bimodality at optical 
bands is an
artifact of their extinction correction which applies to HSB galaxies
alone.  Nonetheless, the fact that the $K^{\prime}$-band is mostly 
insensitive to dust and that the distribution of $\mu_{0,K}^i$ 
is strongly bimodal is sufficient ground for TV97's case of 
structural bimodality in spiral galaxies.

TV97 also considered possible environment dependences of the 
distribution of $\muc^i$ such as effects due to the stripping of gas 
and stars.  They divided their sample into galaxies with and 
without a significant near-neighbour and found, despite the 
statistical limitations, evidence for an enhancement of the 
bimodality in the more isolated galaxies (see TV97 for details). 
Most galaxies with intermediate $\muc^i$ values were found to 
have near significant neighbours.  

Various sources of bias in TV97's analysis should however be
considered before its results can be fully embraced.  These 
include the small size of the UMa cluster sample (Bell \& de Blok 2000), 
the fact that the tentative findings may be particular to the 
UMa cluster (which contains mostly \hi-rich spirals), the need 
for an independent analysis of the data, and the unaccounted 
contribution of the bulge to the disk fits.  The first two issues 
(small-number statistics and UMa-centric analysis) can be addressed 
by repeating TV97's analysis with a larger sample that encompasses 
the full range of Hubble types.  We report such an analysis for 
297 Virgo cluster galaxies in McDonald \etal (2008). 

The need for an independent analysis and the study of the bulge
contribution to the disk scale lengths used by TV97 to infer the 
bimodal distribution of $\muc$ will be addressed simultaneously 
in this paper.  
TV97 obtained $\muc$ from exponential fits that 
avoided the bulge component.
The uncertainty in this ``marking-the-disk'' technique depends 
on the subjective interpretation of the bulge size and the disk 
fit baseline.  For large bulges, a shorter inner fit boundary 
would bias the disk $\muc$ high.  The presence of smaller bulges
should not affect the measurement of $\muc$.

Below, we discuss our own derivation of surface brightness profiles 
from the original deep $K^\prime$-band images from T96 using our 
own isophotal fitting software.  We then decomposed the new 
profiles using our 1D B/D decomposition program.  The measurement 
of $\muc$ by T96 relied on the inward extrapolation of an exponential 
function fitted to the galaxy surface brightness profiles.  We show 
that neglect of the bulge contribution to the light profiles 
has a small and inconsequential effect on the distribution of 
$\muc$ such that bimodality is never erased. 

The outline of the paper is as follows: in \se{data}, 
we briefly discuss our extraction of surface brightness profiles
from isophotal fitting of the calibrated $K^\prime$ images 
originally obtained by T96.  We describe in \se{bddecomp}
the B/D decomposition code that we developed to analyse 
these profiles.  This code relies on the modeling of various 
structural components that are defined in \se{components}
and the fitting procedures are detailed in \se{bdfits}. 
Results are presented in \se{results} as we revisit the 
notion of structural bimodality based on our new data.  
An appendix includes the inferred structural parameters 
for the 65 UMa spiral galaxies. 

\section{Surface Brightness Profiles of UMa Cluster Galaxies}\label{sec:data}

We have reanalyzed the $K^\prime$ photometry for the 65 Ursa Major galaxies 
reported in T96\footnote{The calibrated images are publicly available at 
{\tt http://cadcwww.hia.nrc.ca/astrocat/tully\_help.html}. Please see
T96 for details about the observational set-up.}.  Surface brightness
profiles were extracted using the isophotal fitting method outlined 
in Courteau et al. (1996) and based on the astronomical data reduction 
package 
XVista{\footnote{\tt http://astronomy.nmsu.edu/holtz/xvista/index.html}}. 
The fitted isophotes along elliptical contours have a common center but 
the position angles and ellipticities are variable.  These fits allowed 
for an independent determination of the $K^\prime$-band surface brightness 
profile, contour ellipticity and total $K^\prime$-band magnitude for each 
UMa galaxy.  Our independent analysis of TV97's data will allow us to
check if the reduction methods are a possible source of the 
observed bimodality.

A potential bias could arise through the contribution
of the bulge component to the determination of $\muc$. 
TV97 computed the disk $\muc$ by choosing by eye the radius at which 
the bulge would no longer contribute significantly to the overall profile,
and fitting the outer SB profile, in magnitude units, as a straight line 
from that radius to the end of the profile.  The disk $\muc$ is the inward
extrapolation of that fit to $r=0$.  This marking-the-disk technique
works well on systems with little or no bulge but could potentially
bias the disk $\muc$ high in bulge-dominated systems. The observed 
bimodality could thus be due to systems with dominant bulges unaccounted 
for in the fits by eye\footnote{TV97 were aware of the potential bias
in their fits and attempted to account for it.  They speculated
that the possible bias would not give rise to the observed bimodality.}.

In \se{bddecomp} below, we introduce and make use of a B/D decomposition 
program to separate the bulge light from the disk profile and achieve, 
in principle, more realistic disk fits than TV97. 

\section{Bulge and Disk Decompositions}\label{sec:bddecomp}

The study of the photometric structure of a galaxy can benefit from 
the effective separation of the bulge light from the disk light.  The 
motivation for bulge-to-disk (hereafter B/D) decompositions
originated with de Vaucouleurs (1948), who distinguished between cuspy 
($r^{1/4}$) and exponential surface brightness profiles 
of early to late-type galaxy bulges, respectively.  However, many 
exceptions to the de~Vaucouleurs profile as a description for galaxy 
bulges have since been found (Andredakis et al. 1995; Courteau et
al. 1996; MacArthur et al. 2003; hereafter M03) and it is now generally 
accepted that the projected light profile of a 3D bulge is better 
fit by the generalized \Sersic function:

\be
I(r) = I_0\exp\left[-\left(\frac{r}{r_0}\right)^{1/n}\right]
\label{eq:generalsersic}
\ee
where a \Sersic index of $n=4, 1$, and $0.5$ represents the de~Vaucouleurs,
exponential, and Gaussian profile, respectively.

Fitting the disk portion of the surface brightness profile has,
traditionally, been a more straightforward
task. De Vaucouleurs (1959) found that galaxy disks were best fit
by a simple exponential
function. The exponential distribution of all light and mass density
distributions is indeed ubiquitous in most dynamically relaxed systems
(Lin \& Pringle 1987; Ferguson \& Clarke 2001). However, significant 
deviations from this idealized distribution do exist. Outer truncations,
characterized by a sudden plummeting of the light at large radii, have
been observed in a large fraction of disk galaxies
(M03; Pohlen \& Trujillo 2006). Typically, these
breaks tend to occur in the face-on 
orientation between $\sim$2-3 disk scale lengths and are likely the 
result of disk instabilities (Debattista \etal 2006; Foyle \etal 2008).
Similarly, the inner parts of galaxy disks exhibit departures from 
a pure exponential in galaxies with large bulges (M03). An inner disk 
truncation would explain the observed Freeman Type II galaxies, which 
show a deficit of light at the bulge-disk transition (Freeman 1970; M03). 
Finally, the presence of one or more spiral arms will cause a substantial 
excess of light over the underlying disk which will strongly affect any 
disk fit. Despite these observed deviations from an exponential profile,
galaxy disks are still most often fit by a single pure exponential 
function.

We have implemented a code to model the different components 
of a galaxy based on its 1D azimuthally-averaged surface brightness
profile. Our approach assumes that the galaxy bulge and disk are two
photometrically and dynamically decoupled components and that the disk
can be extrapolated towards the center of the galaxy\footnote{See
Baggett et al. 1998 for B/D modeling that accounts for a putative
inner disk profile break.}. The basic structure of our bulge-to-disk
decomposition code follows from M03 and the majority of our
assumptions and techniques are justified there. The code uses a
non-linear least squares Levenberg-Marquardt technique (Press et
al. 1993) to fit various combinations of functions to a given
surface brightness profile. The bulge and disk light are modelled
simultaneously and the best fit is selected to have the lowest
global $\chi^2$ value. For a thorough discussion of the reliability
and accuracy of this technique, see M03 where the various errors have been
carefully examined via Monte Carlo simulations. We have extended
M03's technique to consider a broader variety of parameterizations for
the bulge and disk light, as well as modelling other components such
as nuclei and spiral arms as we describe below. We do not, however,
model galaxy bars independently since these are not azimuthally
symmetric and thus can have a variety of manifestations in a 1D
profile. However, we will show later that our result is independent of
the exclusion of this component.

Despite the significant insights gained from bulge-disk decompositions 
one must, however, recall the genuine limitations of our simplified
parameterizations.  For instance, the assumption of total decoupling
between the bulge and disk may not be justified. However, there do 
not exist any data or simulations to falsify this hypothesis at
present. Still, the B/D fitting parameters can be complemented 
by non-parametric measurements such as concentrations and effective 
radii, which allow a direct, unbiased comparison of galaxies across
the full Hubble sequence (we explore the distribution of these other
parameters, as applied to the Virgo cluster of galaxies, in McDonald 
\etal 2008). 

\section{Light Profile Modeling}\label{sec:components}
The light profile is a 1D representation of the radial
distribution of light for a given galaxy. 
The shape of
the light profile depends strongly on morphology, from the featureless
exponential disk of LSB spiral galaxies, to the strongly bulged and
spiral arm dominated HSB spiral galaxies, to the nearly pure $r^{1/4}$
profiles of elliptical galaxies. In order to provide a parametric
description of the light profile, various fitting functions are
needed. Parameterizations for the nucleus, bulge, disk, spiral arms
and disk truncation are discussed below.

\subsection{Nucleus}
It has become clear that most galaxies harbour either a bright
compact nucleus in their center (B{\"o}ker et al. 2002)
or a supermassive black hole (SBH), or both (C\^ot\'e et al. 2006). 
The typical half-light radius of a nucleus ($<$15pc; C\^ot\'e \etal 
2006) would cover $\sim$1-2 pixels on most ground-based detector 
with modest ($\sim$ 0.25\arcsec/pix) resolution.  These compact 
objects can be treated as a point source whose apparent width is 
determined by the ambient seeing.  Compact nuclei differ significantly 
from the galaxy bulge both in brightness and extent. Nuclei are typically 
100 times smaller than galaxy bulges and contribute, on average, 0.3\%, 
and sometimes as much as 10\%, of the total galaxy luminosity 
(C\^ot\'e et al. 2006).  For comparison, we find that the bulge 
of a typical Sb-Sc galaxy contributes roughly 10\% of its total 
luminosity. We will treat galaxy nuclei as point sources, described 
by a delta function convolved with a Moffat function to simulate the 
effect of atmospheric turbulence (seeing). 

\begin{figure*}
\centering
\begin{tabular}{cc}
\includegraphics[width=0.45\textwidth] {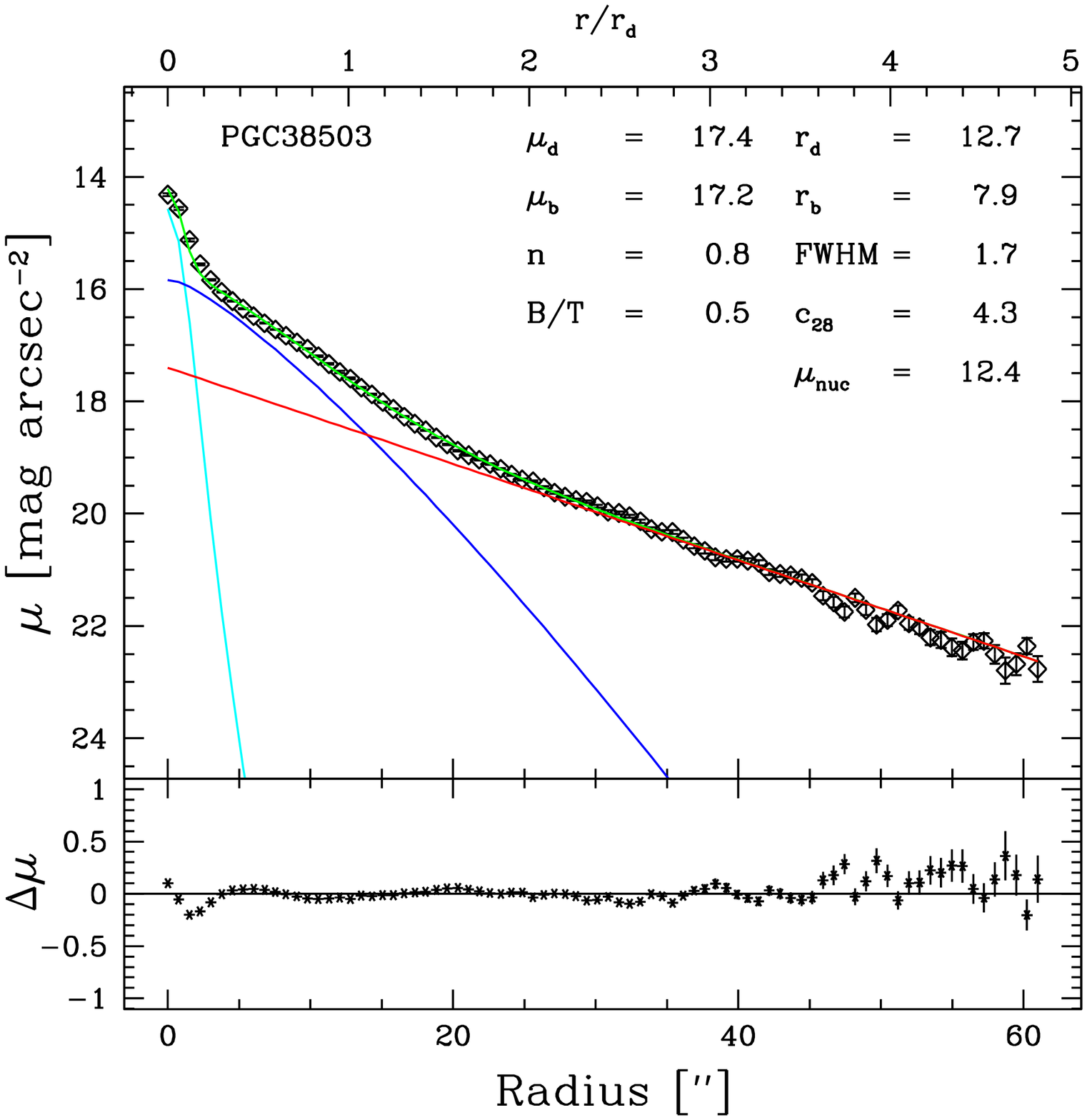} &
\includegraphics[width=0.45\textwidth] {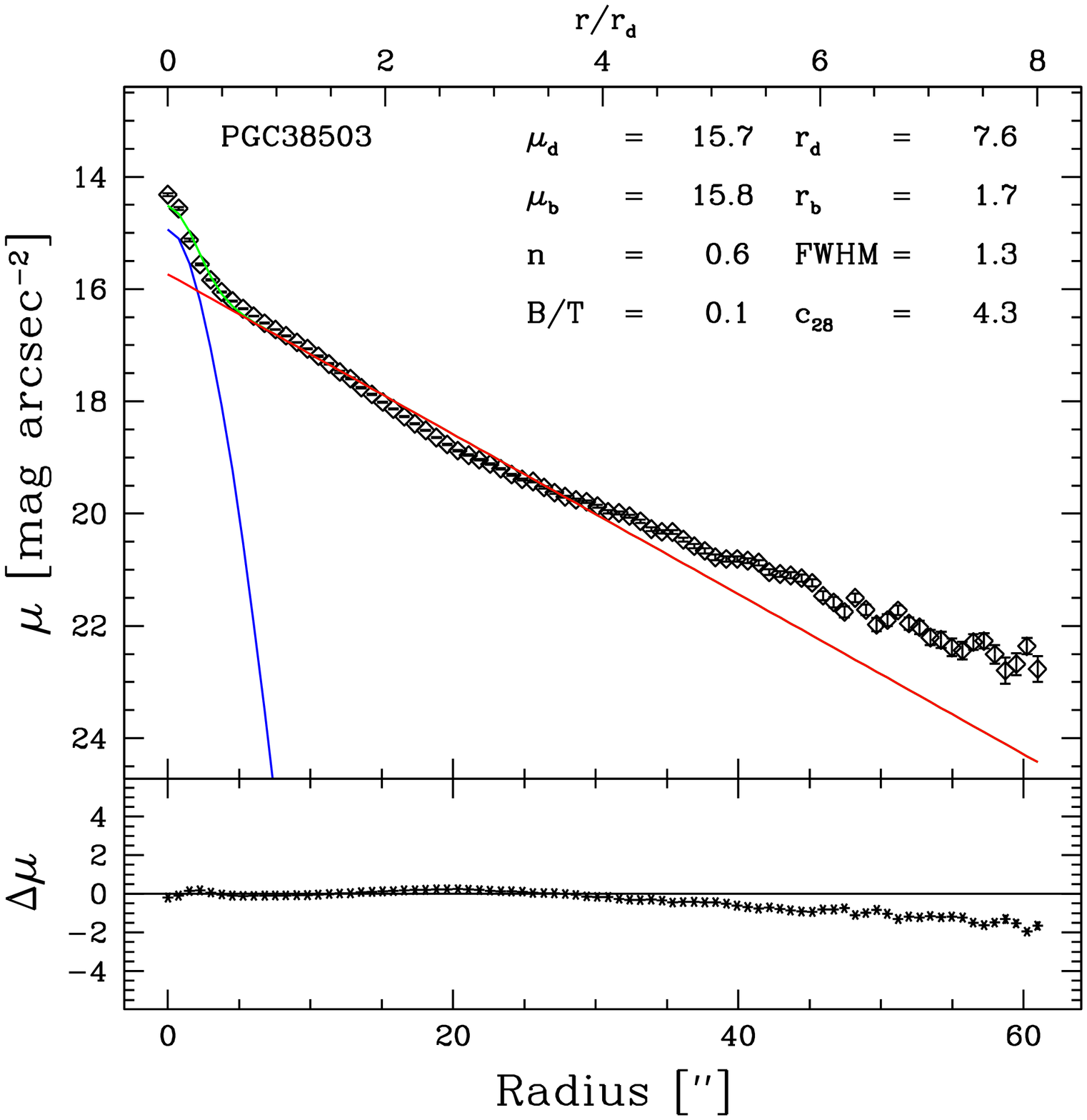} \\
(a) & (b) \\
\end{tabular}
\caption{B/D decompositions of $K^\prime$-band surface brightness profile with
  a central nucleus (a) and without (b). The justification for a
  nucleus component seems obvious in this case. The red line is the
  disk (exponential) fit, the blue line is the bulge (S\'ersic) fit
  and the cyan line is the fit to the nucleus. The
  decomposition parameters all refer to the $K^\prime$-band.}
\label{fig:nucleus}
\end{figure*}

\Fig{nucleus} shows the effect of modelling a nucleus component
on the surface brightness profile of a sample galaxy. This figure
depicts a bulge-disk decomposition for a galaxy with an obvious
nuclear component. If we allow an extra free parameter for the
nucleus, a much better fit to all three components (nucleus,
bulge, disk) is obtained. Without the fit to the nucleus, the
bulge fit is forced to be cuspy, in order to account for this extra
light at the center. 

Note the dramatic effect of this extra component on the overall 
bulge-disk decomposition.  A nucleus is only fitted if the residuals 
(lower panels in \Fig{nucleus}) in the very central parts of the galaxy 
are strongly negative.

\subsection{Bulge}

The bulge component of each galaxy is modelled as a generalized \Sersic
(1968) function:

\be
I_b(r) = I_e\exp\left\{-b_n\left[\left(\frac{r}{r_e}\right)^{1/n}-1\right]\right\},
\label{eq:sersic}
\ee
or, in magnitudes,
\be
\mu_b(r)=\mu_e+1.0857~(b_n)\left[\left(\frac{r}{r_e}\right)^{1/n}-1\right]
\ee

where $\re$ is the radius which contains 50\% of the total galaxy
luminosity, or the effective radius. The total galaxy luminosity is 
determined by extrapolating the profile fit from $r=r_{max} \rightarrow \infty$ 
and adding this extrapolation to the total 
light within $r_{max}$.  The effective surface brightness, $I_e$, is defined 
as the surface brightness at the effective radius.  The parameter $b_n$ ensures 
that $\re$ is the half-light radius. This parameter cannot be solved for 
analytically and we use the numerical approximations of Ciotti \&
Bertin (1999) and M03. The \Sersic index $n$ allows for the fitting of dominant 
(e.g. de Vaucouleurs, $n=4$) classical bulges, as are observed in massive 
galaxies, as well as nearly exponential ($n=1$) bulges in late-type spirals. 
By allowing the \Sersic index to vary from 0.1 to 10.0, we can fit a wide 
range of bulge shapes from LSB galaxies with $n\lesssim 0.5$ bulges to the 
most massive galaxies with $n\gtrsim 6$. The typical error in $n$ is $\sim 0.2$.

\subsection{Disk}
 
The radial density profile of a spiral galaxy disk is generally
described by an exponential function of the form:

\be
I_d(r) = I_0\exp\left(-\frac{r}{h}\right),
\label{eq:diskintensity}
\ee
or, in magnitudes,

\be
\mu_d(r) = \mu_0 + 1.0857\left(\frac{r}{h}\right),
\label{eq:expdisk}
\ee
where $I_0$ and $\muc$ are the disk central surface brightness in
intensity and magnitude units respectively, and $h$ is the disk scale
length.  Eqs. \ref{eq:sersic} \& \ref{eq:diskintensity} are equivalent for
$n=1$. The exponential function works well for galaxies with little or
no radial structure like LSB disk galaxies.  However, non-axisymmetric
structures such as bars or
spiral arms, and inner or outer profile truncations will induce
departures from a pure exponential disk fit. Our modeling tries to
account for these features. For disks with noticeable departures from
the pure exponential model, the \Sersic function (\Eq{sersic}) 
is a better description. 

\begin{figure*}
\centering
\begin{tabular}{cc}
\includegraphics[width=0.45\textwidth] {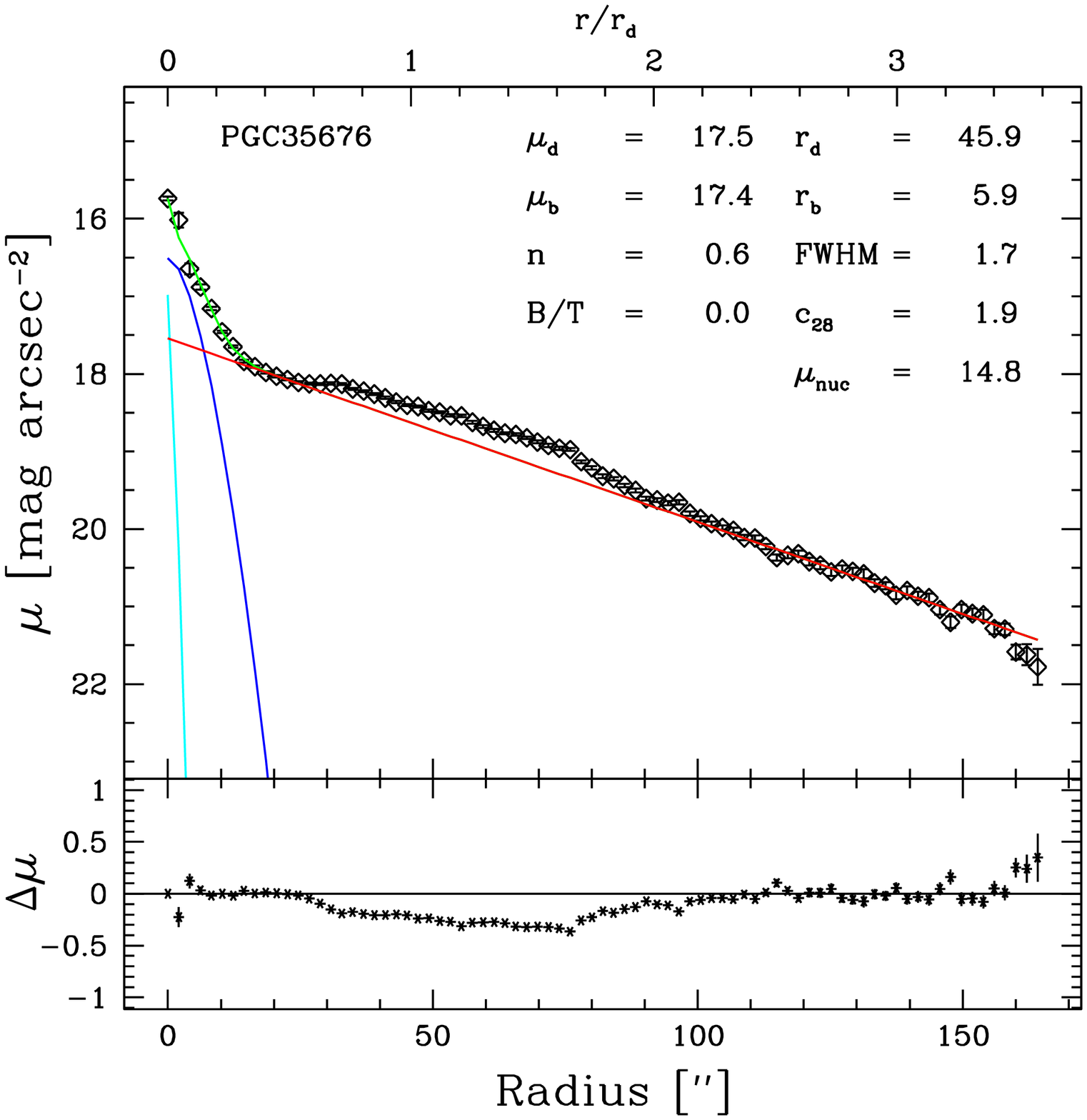} &
\includegraphics[width=0.45\textwidth] {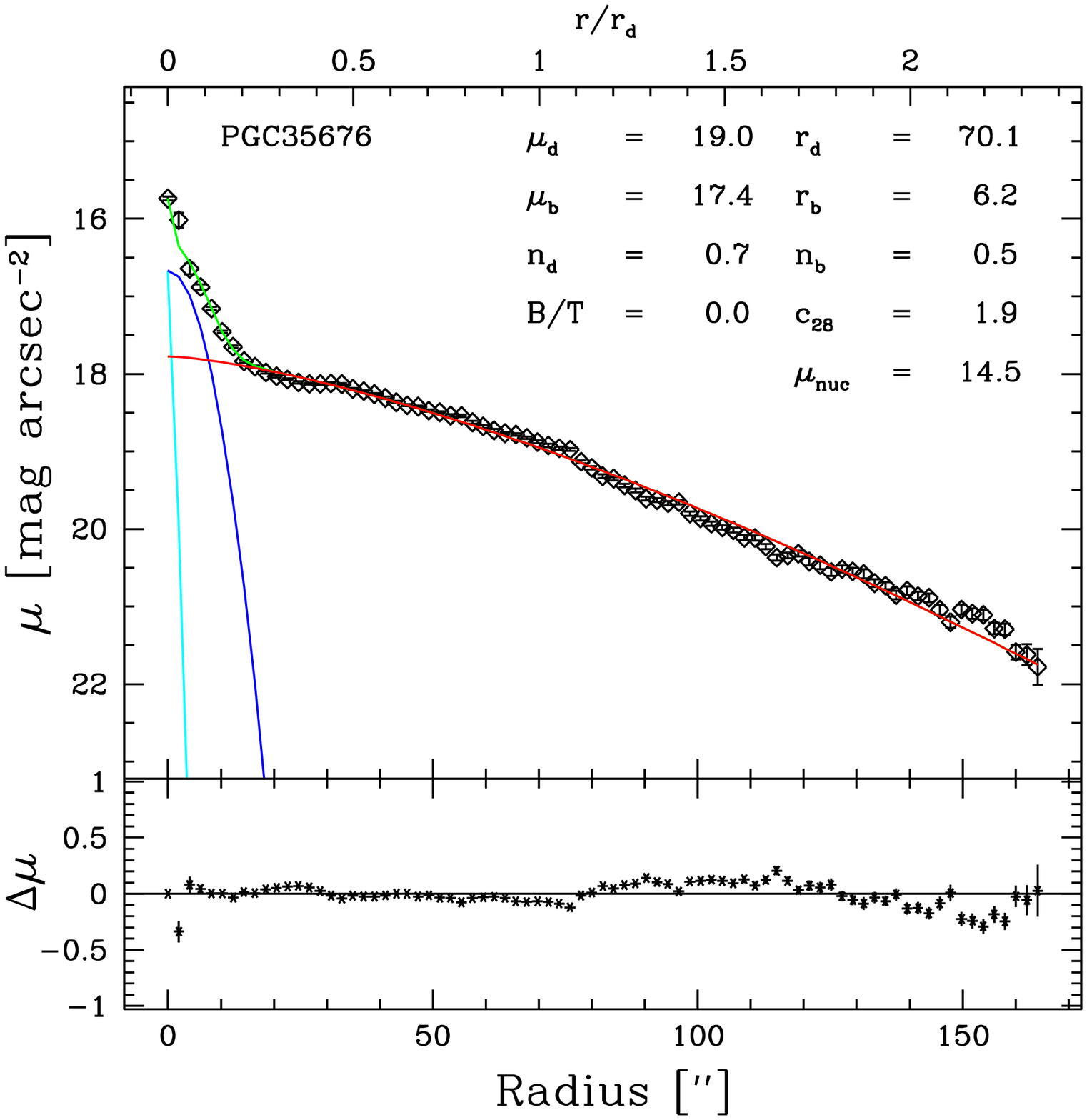} \\
(a) & (b) \\
\end{tabular}
\caption{Bulge-disk decompositions with an exponential (a) and S\'ersic (b)
  disk. The decomposition all parameters refer to the $K^\prime$-band.}
\label{fig:highlow-n}
\end{figure*}

High-$n$ disk fits can describe the so-called Type III interacting 
galaxies (Erwin et al. 2005), while $n<1$ fits can describe cored 
disks, found typically in dwarf systems. \Fig{highlow-n} shows
how a \Sersic disk can yield a more compact bulge.

\subsection{Spiral Arms and Rings}

\begin{figure*}
\centering
\begin{tabular}{cc}
\includegraphics[width=0.45\textwidth] {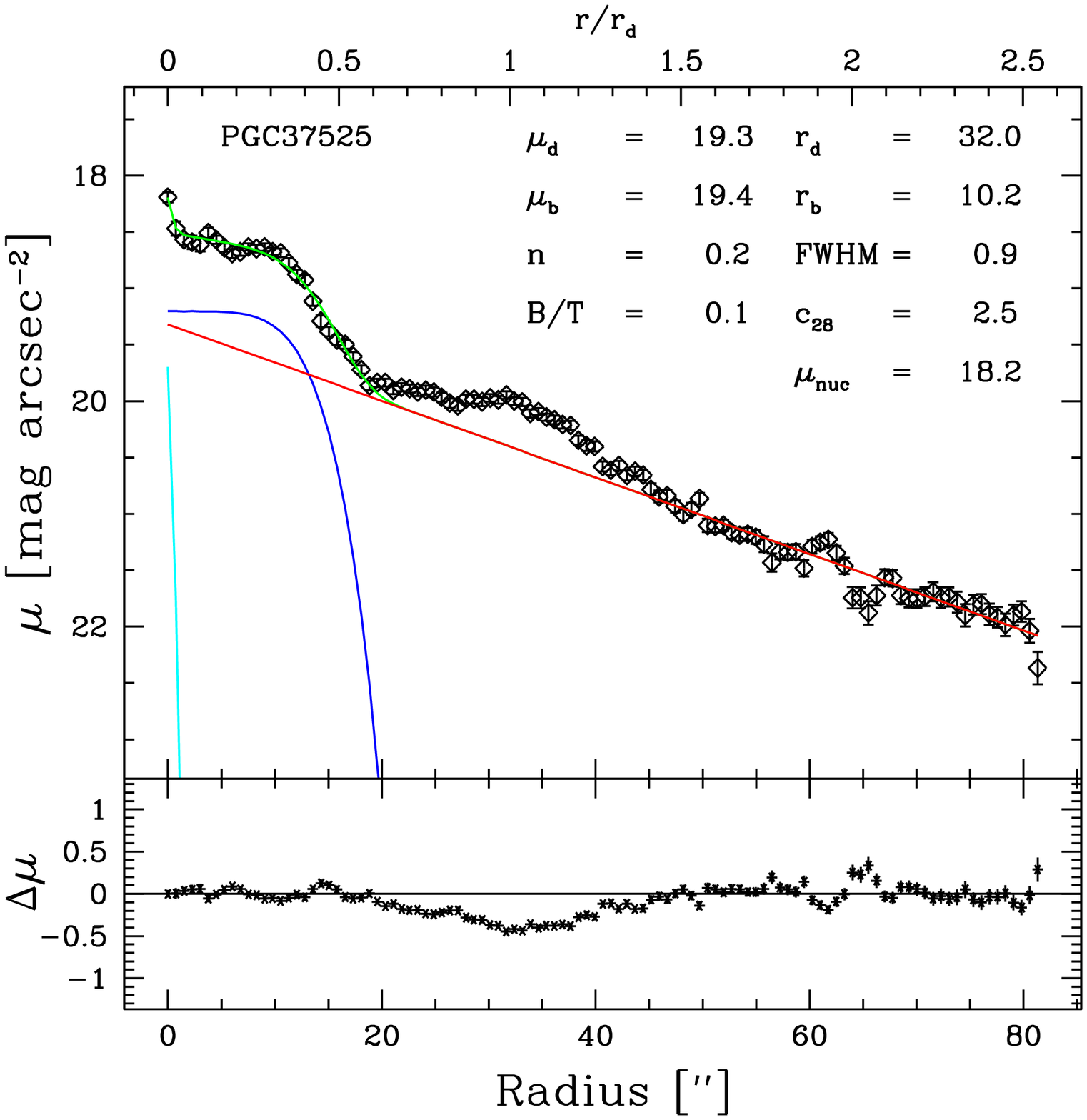} &
\includegraphics[width=0.45\textwidth] {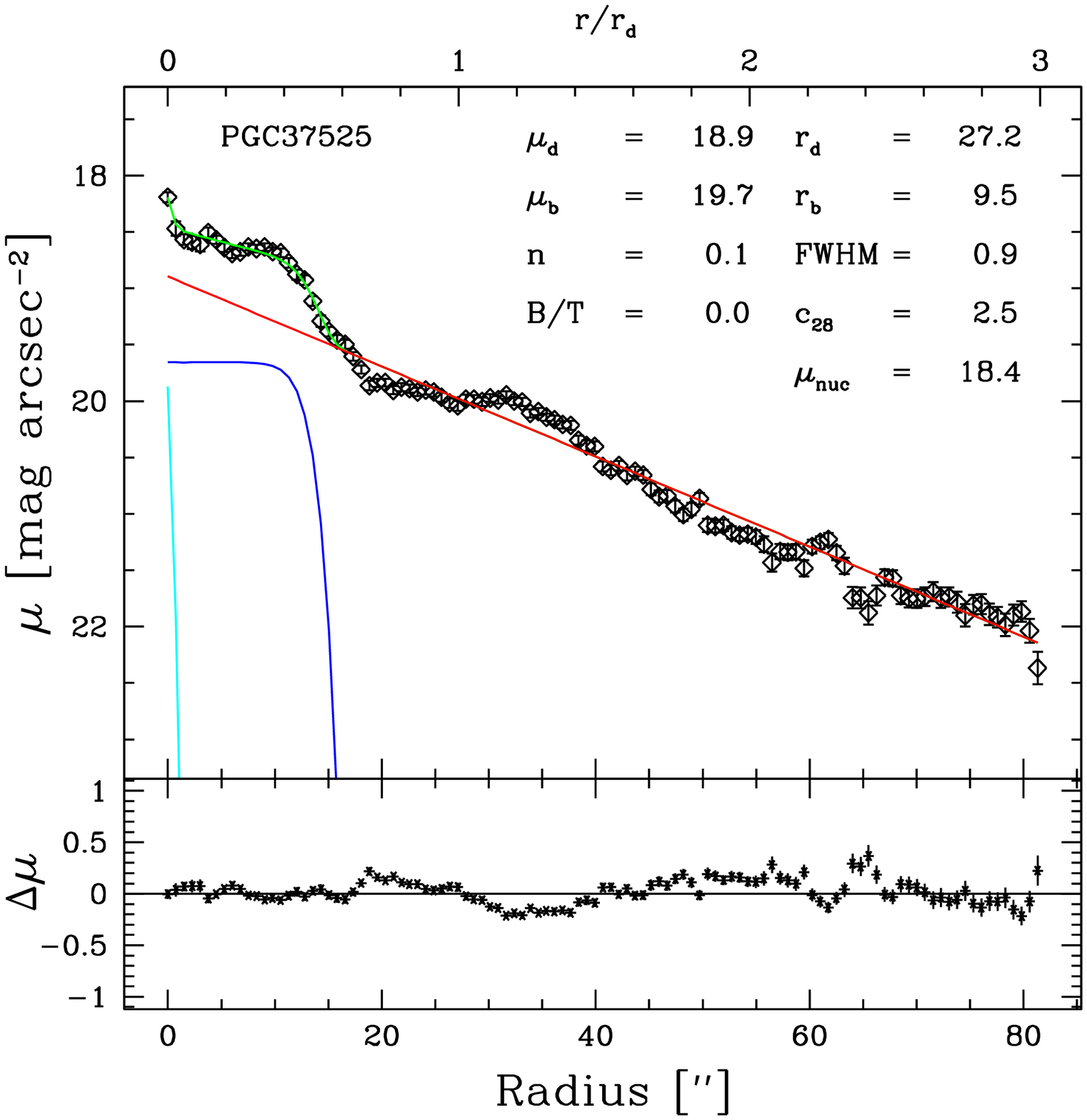} \\
(a) & (b) \\
\end{tabular}
\caption{Bulge-disk decompositions with (a) and without (b) spiral arm
  clipping. The decomposition parameters all refer to
  the $K^\prime$-band.}
\label{fig:spiral}
\end{figure*}

In azimuthally-averaged surface brightness profiles, both rings and
spiral arms generally manifest themselves as smooth fluctuations above the
underlying exponential disk (radial profile cuts would show more
abrupt features). Assuming that rings and spiral arms represent an excess of
light on top of the underlying disk, we mostly wish to trace the
underlying exponential profile and clip any excess.  We identify 
the starting point of any arm/ring by finding the locations where the 
surface brightness profile deviates positively from the mean 
exponential fit.  The end point of the arm/ring will be the location 
where the surface brightness profile returns to the exponential 
fit value. If the isolated arm/ring region is found to have a smooth 
convex shape about the disk and lies beyond one disk scale length, 
then that region is excluded from the disk fit. The bulge and disk 
would be fit once more, with the clipped regions being excluded 
from the fit.  This procedure is iterated until the extent of 
the arm/ring remains constant. \Fig{spiral} shows an example of this 
process.  In addition to yielding a better disk fit, the exclusion 
of spiral arms in the B/D decomposition has a non-negligible 
effect on other galaxy components.  It is clear from \Fig{spiral} 
that the nucleus and bulge fits are significantly altered by 
the removal of a spiral arm.

\subsection{Outer Truncations}

\begin{figure*}
\centering
\includegraphics[width=0.9\textwidth] {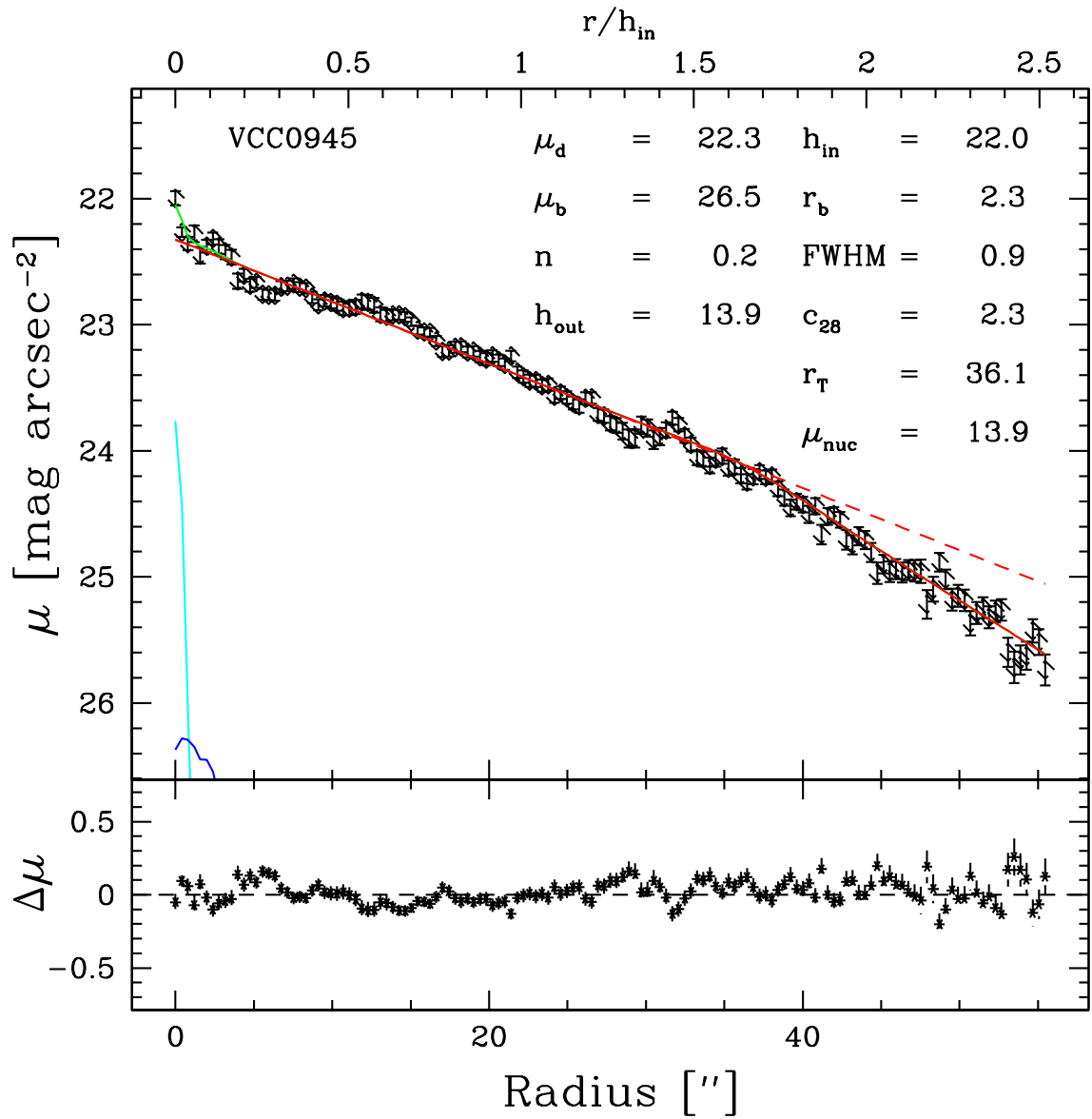}
\caption{Example of a B/D decomposition for a Virgo cluster galaxy with an outer
  truncation. The two-component disk fit with two exponentials is
  governed by a simple step function. The decomposition parameters all
  refer to the $r$-band.}
\label{fig:truncation}
\end{figure*}

A topic of great interest in galaxy structure studies is that 
of disk truncations (Pohlen \& Trujillo 2006; Ro{\v s}kar et al. 2008;
Foyle et al. 2008). While the cause of these truncations remains
unclear, their signatures as departures from the extrapolated 
inner disk to greater radii are easily identified, especially 
in edge-on galaxies.  Because truncations may be mistaken for, 
or induced by, sky subtraction errors, care must be taken when 
measuring sky levels.  An exact definition of disk truncation
is still missing, however we choose to model them as the transition
between an inner and outer exponential profiles governed by 
a Heaviside step function, $S(r)$, as in:

\be
I_d(r) = I_{0,in}\exp\left\{\frac{-r}{h_{in}}\right\}\times(1-S) + I_{0,out}\exp\left\{\frac{-r}{h_{out}}\right\}\times S,
\ee
where 
\be
S = {\frac{1}{[1+\exp(r_t-r)]}}.
\ee
For $r \ll r_t$, S $\rightarrow$ 0 and the inner component
dominates. For $r \gg r_t$, S $\rightarrow$ 1 and the outer component
dominates. This choice of fitting function requires that two additional
parameters be defined; an outer disk scale length, $h_{out}$, and a
truncation radius, $r_t$. The truncation radius is defined here as the
radius where the disk fit is comprised of equal parts of each of the
two exponential components. The parameter, $I_{0,out}$, is fixed such
that the two disk components intercept at $r_t$.  
An example of B/D fitting with a profile break is shown in 
\Fig{truncation}.  

In our sample of 65 UMa galaxies, none have clearly defined
truncations. However, although unused in this particular study,
the ability to fit these peculiar cases is an important functionality
of any bulge-disk decomposition code and will be more relevant in a
future application to the richer and more exotic Virgo cluster
(McDonald \etal 2008). 

\subsection{Seeing}
\begin{figure*}
\centering
\includegraphics[width=0.9\textwidth] {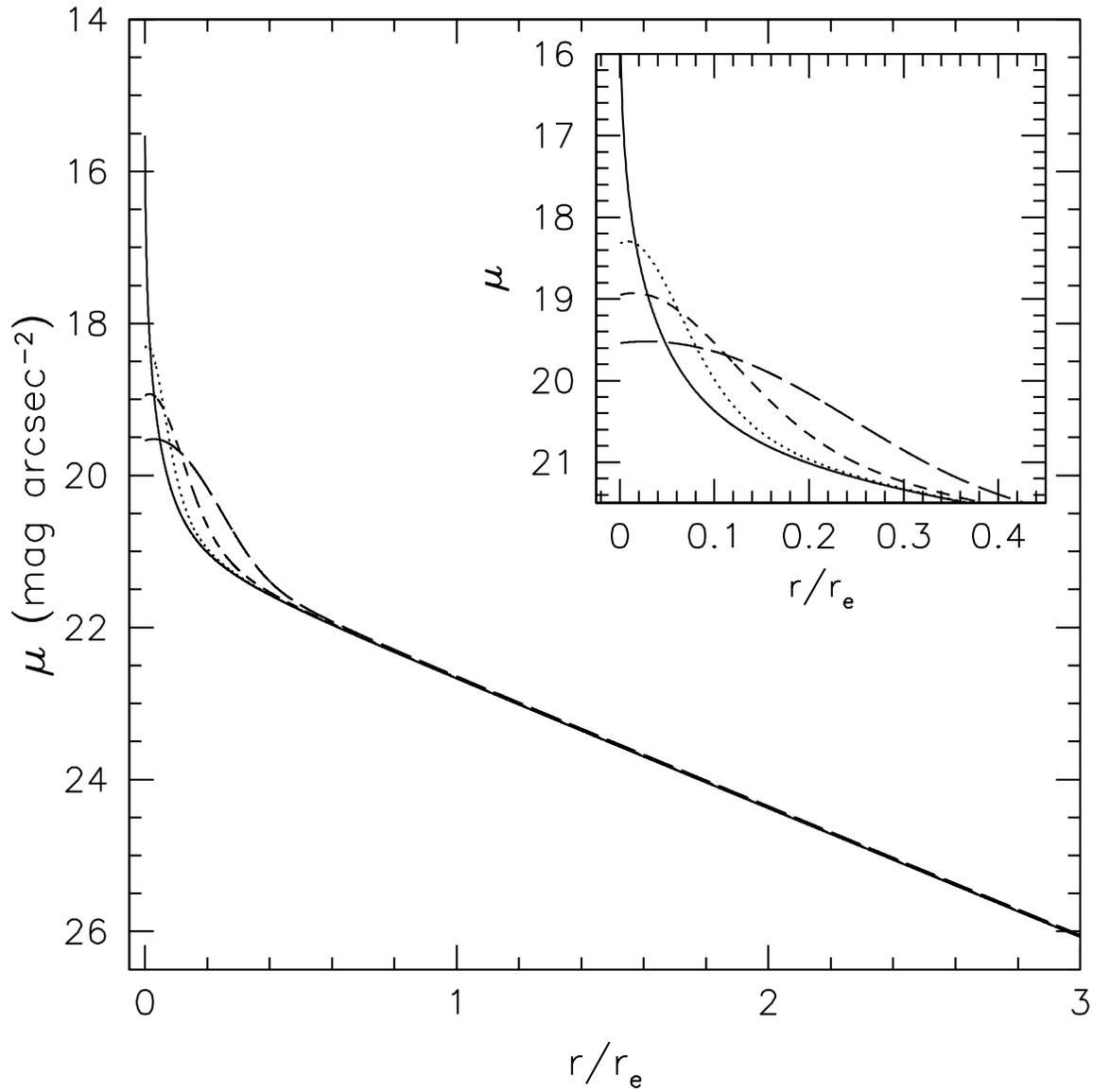}
\caption{Example of a model $n=4$ S\'ersic bulge and exponential disk
  (solid) convolved with a Moffat function of FWHM=2$''$ (dotted),
  FWHM=4$''$ (short dash) and FWHM=8$''$ (long dash). The inset shows
  a magnified portion of the inner profile.}
\label{fig:convolution}
\end{figure*}

Sharp light profile features, such as a galaxy nucleus 
or a cuspy bulge, are naturally blurred by the atmosphere. 
In order to model this effect, the bulge and nucleus fitting
functions are convolved with a Moffat (1969) function whose FWHM 
matches the average seeing for each observation. The Moffat 
function is defined as:

\be
I_{_{\rm PSF}}(r)={\frac{\beta-1}{\pi\alpha^2}}\left[1+\left(\frac{r}{\alpha}\right)^2\right]^{-\beta}
\ee
where $\beta$, the shape parameter, is taken to be 4.765 (Trujillo et
al. 2001) and $\alpha$ is related to the FWHM via 
FWHM$=2\alpha\sqrt{2^{1/\beta}-1}$. The convolution of a point source
with the seeing function yields satisfactory fits to the nucleus of
our sample galaxies, as is seen in \Fig{nucleus}. The effect of
numerical convolution on a \Sersic profile with $n=4$ is shown in
\Fig{convolution} and its inset. The effect of seeing for any
profile with $n\gtrsim$2 is indeed significant and seeing
measurements, for the sake of bulge profile determination, should be
determined carefully. Fortunately, the quantity that matters most for
this study, the disk $\muc$, is only weakly affected by seeing. 
For computational efficiency, the convolution is only applied 
out to the radius where the fractional difference between the
convolved and unconvolved profiles is $<$0.1\%. 

\section{Fitting Procedure}\label{sec:bdfits}
In order to converge upon a single decomposition for a given 
surface brightness profile, and having determined the types of 
galaxy components to fit, we now resort to a combination of grid
search and generalized least-squares fitting to identify the best 
fitting functions.  In the following sections the procedure 
from surface brightness profile to final decomposition is
outlined in detail, inspired greatly by M03.

\subsection{Initial Estimates}
Non-linear least-squares fitting requires a set of initial guesses
for the model parameters (Press et al. 1993). 
We can make use of various approximations guided by previous 
similar efforts (e.g. M03) to guess the best shape of a light 
profile with minimal computational effort; these are described 
below. For each parameter, several initial estimates are 
identified in order to achieve the best possible
sampling of the $\chi^2$ manifold.

\subsubsection{Exponential Disk}
The disk central surface brightness, $\muc$, and scale length, $h$,
are first estimated with a marking-the-disk technique
(Giovanelli et al. 1994; Courteau et al. 1996; TV97). This
technique consists of fitting the surface brightness profile 
in magnitude scale with a straight line over a prescribed baseline.  
We adopt for our fits the radial range $r_e<r<r_{max}$, 
where $r_e$ is the half-light radius of the total extrapolated light profile and $r_{max}$ is the last reliable data point (with surface 
brightness error $<$ 0.15 mag arcsec$^{-2}$), yielding the initial 
parameters $\muc$ and $h$.  This assumes that the outer profile 
at and beyond $r_e$ is dominated by the disk component. 

\subsubsection{\Sersic Disk}
The initial estimates for a \Sersic disk are the same as those used for
the pure exponential disk. For a pure exponential disk,
the following conversion applies:

\be
\mu_e=\mu_0+1.822,
\label{eq:e-0}
\ee
\be
r_e=1.678h.
\ee
These relationships arise from matching equations Eqs. \ref{eq:sersic}
\& \ref{eq:diskintensity} with $n=1$ and $b_{n=1}=1.678$.

\subsubsection{Truncation Radius}
The analysis of surface brightness profiles for face-on galaxies and
profile cuts for edge-on galaxies have yielded a broad range in
truncation radii, $r_t$, from 2.5 to 4.5 disk scale lengths (Pohlen \&
Trujillo 2006).  
Our general procedure uses, as first guess, $r_t\sim$0.5$r_{max}$ 
and $r_t\sim$0.75$r_{max}$. We find that our decompositions 
converge upon an acceptable truncation radius, as gauged by eye, 
when applicable.  Recall that none of the UMa galaxies in this 
study show a significant disk truncation. 

\subsubsection{\Sersic Bulge} 
M03 identified 3 sets of initial parameters for galaxy bulges. The
first set is determined empirically from observations of late-type
field spirals (Courteau et al. 1996): $\mu_e=\muc$ and $r_e=0.15h$, 
where $\muc$ and $h$ are measured from the marking-the-disk technique. 
We find exactly the same result for UMa galaxies in \se{results}. The
second set of initial parameters is obtained by first subtracting the
marking-the-disk fit from the surface brightness profile. In an ideal
case, the only remaining light would be that of the bulge and
possibly a nucleus. The bulge effective surface brightness and radius
are then measured directly from the residual profile, yielding the
second set of initial bulge parameters. The third set of guesses comes
from scaling $r_e$ and $\mu_e$ to different values of the \Sersic $n$
parameter (M03): $r_e=(b_n/0.4343)0.15h$ and
$\mu_e=(b_n-b_{n=1})+\mu_0$.

\subsection{Grid Search}
We have so far identified 5 types of profile fits: 
$(i)$ single exponential,
$(ii)$ single \Sersic, 
$(iii)$ \Sersic bulge plus exponential disk, 
$(iv)$ \Sersic bulge plus \Sersic disk, and 
$(v)$ \Sersic bulge plus a 2-component exponential disk. 
For each profile decomposition, our grid search involves 3 sets of initial
parameters, as well as 3 different values for the seeing FWHM (the
mean value and the high/low values corresponding to typical
measurement errors of 15\%). Finally, 100 different search values for
the \Sersic $n$ parameter were used for the bulge (0.1-10.0) and 20 for
the disk (0.1-2.0), both in $\Delta n=0.1$ increments. These values
were held fixed during the fitting process in order to provide a more
stable solution (M03). This set of parameters required 900 different
fits for decompositions with a \Sersic bulge and one or two exponential
disks and 6000 fits for decompositions with a \Sersic bulge and a
\Sersic disk. 

\subsection{Fit Selection}
In order to achieve a unique, best model solution, one must devise 
a scheme to discard unwanted fits. To narrow down our suite of hundreds
or thousands of fits, we apply various rules in several stages which
eventually yield a single fit. The first pass discards any fits that
have not converged to a final solution after 50 iterations.  This
requirement ensures that any final solution is indeed a local minimum
in the $\chi^2$ space. Other constraints devised by M03 apply mostly
to late-type spirals; our software only considers them in the case 
of a \Sersic bulge and exponential disk. These are:
\newline
\vbox{
\hbox{1. $r_e > (0.3)^{1/n}(FWHM)$ }
\hbox{2. $B/D < 5$ }
\hbox{3. $r_e/h < 1$}
\hbox{4. $h < r_{max}$.}
}
\newline
The first constraint requires that the size of the bulge be larger than
the seeing disk; the second requires a reasonable B/D ratio (determined
empirically for late-type galaxies); the third requires that the disk 
extends further than the bulge; and the fourth requires that the profile
extends to at least one disk scale length.

Selection from
the remaining fits is on the basis of both their global
and inner $\chi^2$ values, the latter being the $\chi^2$ evaluated
within half of a galaxy's effective radius (M03). This process
involves sorting the remaining fits in terms of their {\it{inner}}
$\chi^2$, removing the lower half (worst solutions), then sorting the
remaining fits by their {\it{global}} $\chi^2$ and again removing the
lower half. This process is iterated until less than 50 fits
remained. At this point, the remaining fits are ranked based on both
the global and inner $\chi^2$ values and the fit that has the lowest
total position on both lists is chosen. This process is applied to
each of the 5 basic types of fits as described in \S5.2. The best
decomposition is the one with the lowest global $\chi^2$ per degree of
freedom. The global, rather than the inner, $\chi^2$ is preferred here
since the 5 different fits differ most strongly in their treatment of
the outer profile.

\section{UMa Results}\label{sec:results}
\begin{figure*}
\centering
\begin{tabular}{cc}
\includegraphics[width=0.45\textwidth] {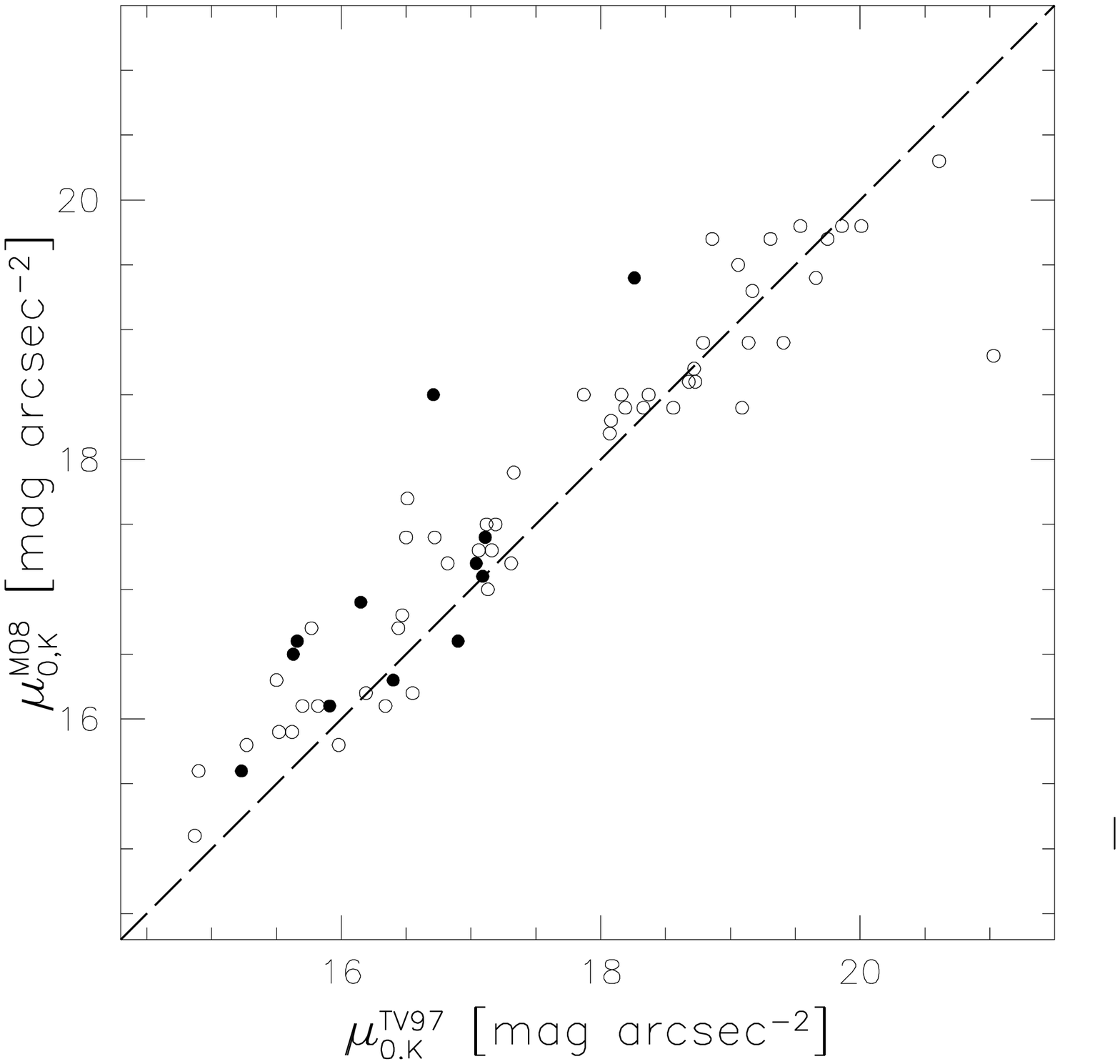} &
\includegraphics[width=0.45\textwidth] {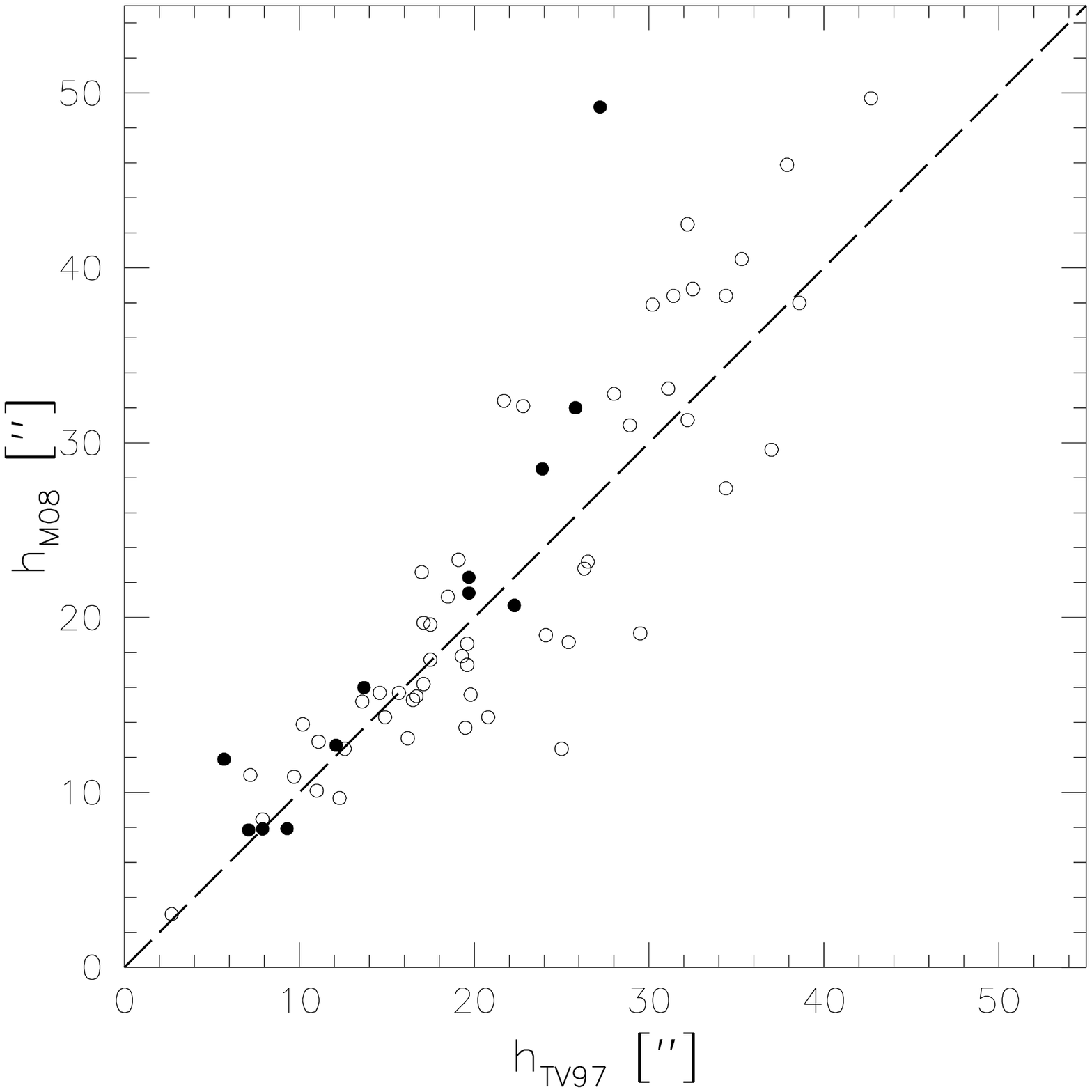} \\
(a) & (b) \\
\end{tabular}
\caption{Comparison of measured (a) $\mu_0$ and (b) $h$ as published
  by TV97 and from our bulge-disk decompositions. The open circles are
galaxies with bulge-to-total (B/T) ratios of less than 20\%, while the
filled circles have B/T greater than 20\%.}
\label{fig:tvcompare}
\end{figure*}

The results of the modeling of 65 UMa profiles with a Sersic bulge and an exponential disk using the methodology described above, are presented in Table 1.
The values 
of the disk $\muc$ were corrected for inclination and extinction according to 
\Eq{inc_corr}, assuming the transparent case ($C^K=1$). 
The broad effect of performing a B/D decomposition should be to 
lower the measured disk $\muc$ since we are now subtracting the
bulge light from the profile. 
Interestingly, our B/D decomposition of the 65 $K^\prime$-band UMa 
surface brightness profiles reveals results very similar
to those obtained by TV97. Indeed, \Fig{tvcompare} shows the
correlation between $\mu_0$ and $h$ as measured by TV97 and
ourselves. Outliers with high bulge-to-total ratios are expected since
TV97 did not parametrize the bulge and thus were unable to fully
subtract the light it contributed to the disk. LSB outliers are likely
due to disks with non-exponential profiles which the eye will fit very
differently than an automated process.

\begin{figure*} 
\centering
\includegraphics[width=0.9\textwidth] {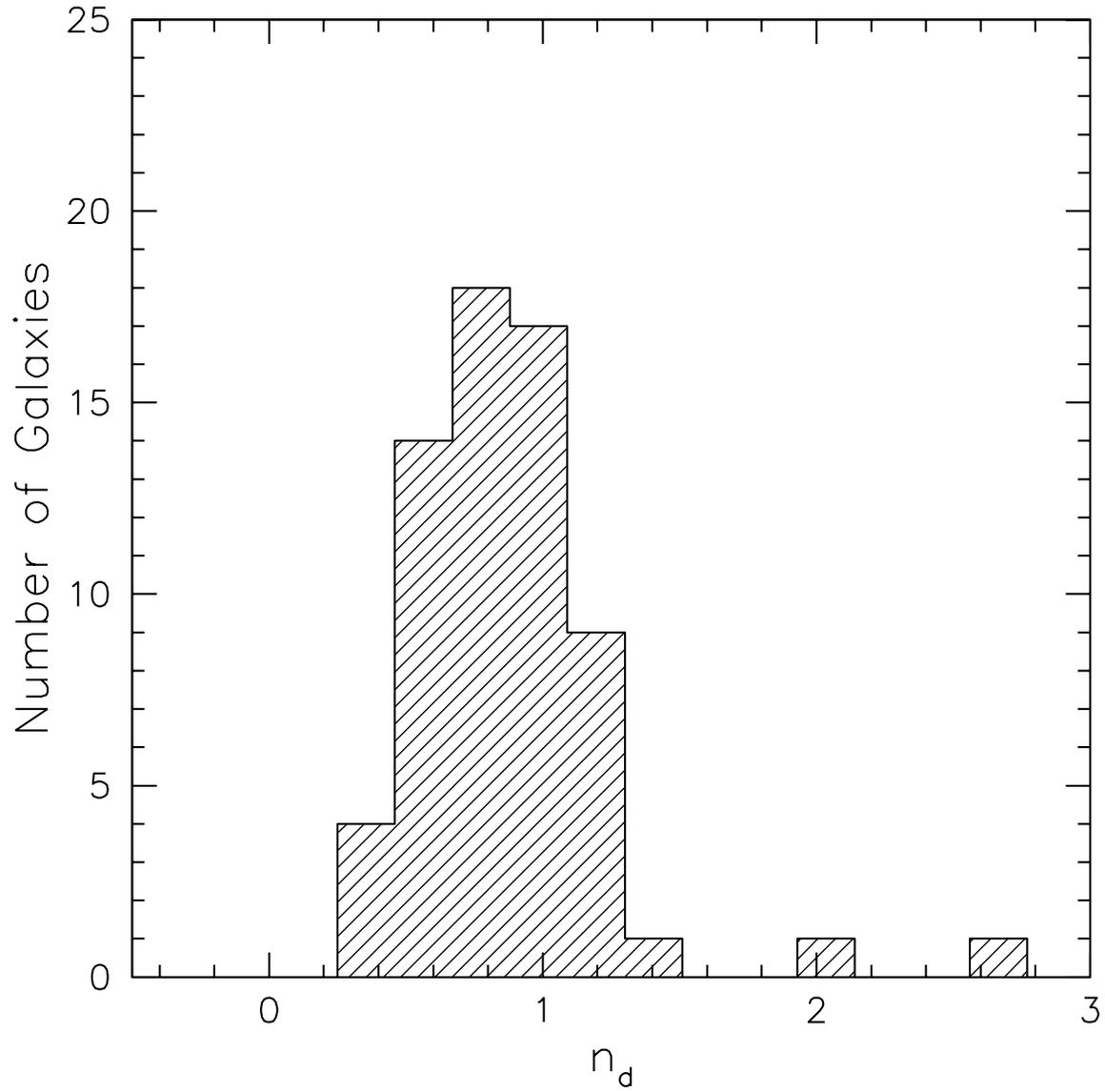}
\caption{Distribution of S\'ersic n parameters for the disks of the sample of 65
  UMa galaxies computed from $K^\prime$-band surface brightness profiles..}
\label{fig:ndist}
\end{figure*}

Before determining the distribution of disk central surface
brightnesses, we can verify the adequacy of the exponential function
to characterize galaxy disks. \Fig{ndist} shows the distribution of 
the \Sersic $n$ parameter for the extended component of galaxies in our
sample. The peak of this distribution around $n=1.0$ suggests that the
exponential function is generally a suitable fitting function for
galaxy disks. However, there exists galaxies with high-n values
(S0s) and with low-n values (typically irregular or dwarfs)
suggesting that parameterizing all disks with the same function
may be hazardous. The two high-$n$ outliers in \Fig{ndist} correspond
to profiles which have a smooth transition between bulge and disk and
thus were best fit by a single \Sersic function, yielding a cuspy
profile. We return to this topic at the end of the 
section by using a \Sersic function rather than the exponential 
to derive the distribution of disk surface brightnesses. 

\begin{figure*} 
\centering
\includegraphics[width=0.9\textwidth] {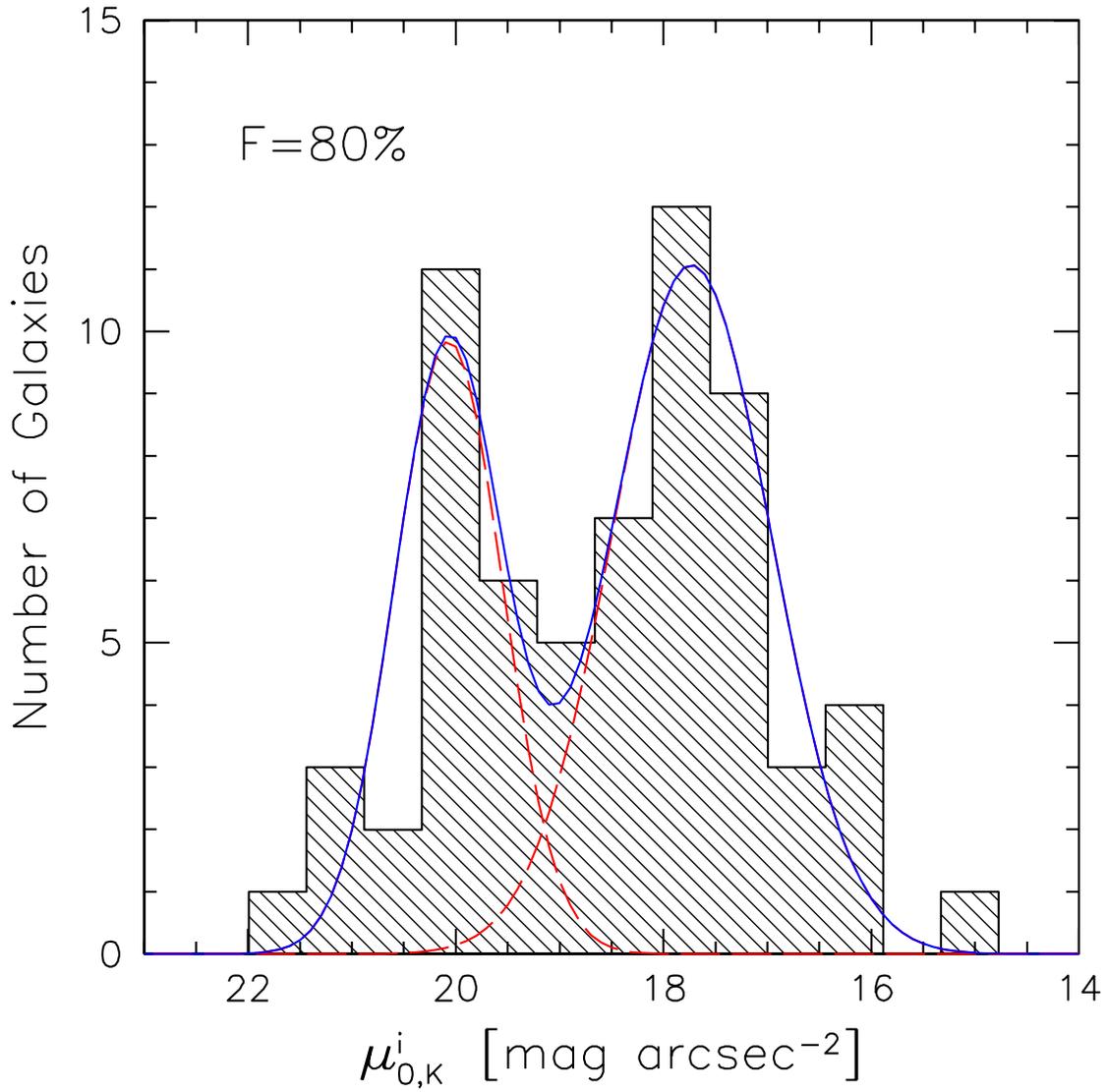}
\caption{Distribution of inclination-corrected $K^\prime$-band disk central
  surface brightnesses, $\mu_{0,K}^i$ for the sample of 65 UMa galaxies. 
  The solid line represents the sum of the two Gaussian fits 
  (dashed) to the histogram. The F-test confirms with 80\% probability that this distribution for $\mu_{0,K}^i$ cannot be described by a Gaussian function.}
\label{fig:uma_csb}
\end{figure*}

The distribution of inclination-corrected disk $\mu_{0,K}^i$, 
shown in \Fig{uma_csb} for 65 UMa galaxies is clearly bimodal with
peaks, as fit by two Gaussians, at 20.0 and 17.8 $K^\prime$ \magarc. 
These values can be compared with TV97's reported peaks
at 19.7 and 17.3 $K^\prime$ mag arcsec$^{-2}$. 
Within the errors, the location
of our and TV97's peaks are the same.  
This excellent agreement between the marking-the-disk
technique and B/D decompositions has been shown in previous studies
(e.g. de Jong 1996). However, this represents the first test of this
kind using TV97's basic data, confirming the adequacy of their analysis. 

In order to gauge the strength of the bimodality, we use the
statistical F test which compares two different fitting functions 
for a single set of data. We find that the
distribution of disk central surface brightnesses 
for the 65 galaxies in this sample is better 
fit with a pair of Gaussians, rather than a single Gaussian, 
with 80\% confidence. If we remove galaxies that do not exhibit a
clear exponential disk (e.g. S0s) from this sample, the confidence
level increases to 85\%.

\begin{figure*}
\centering
\includegraphics[width=0.9\textwidth] {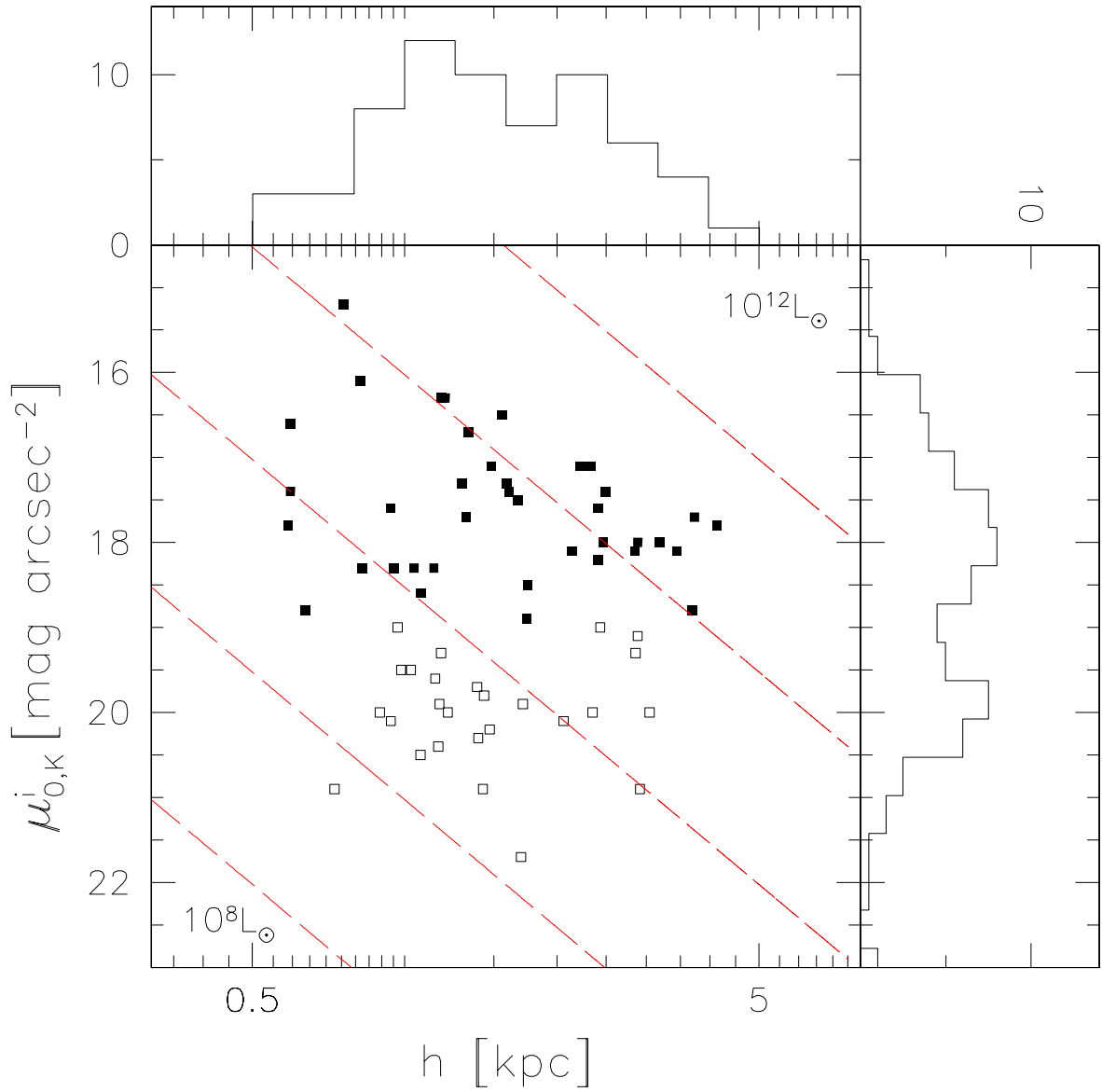}
\caption{Distribution of inclination-corrected $K^\prime$-band disk central 
surface brightnesses, $\mu_{0,K}^i$, as a function of disk 
scale length, $h$, measured in kpc (using a distance to UMa of 15.5 
Mpc) for the sample of 65 UMa galaxies. The long-dashed lines represent lines of constant
total disk luminosity while the different surface brightness classes are
identified by open squares (LSB) and filled squares (HSB).}
\label{fig:csb_h}
\end{figure*}

\Fig{csb_h} shows the relationship between $\muc$ and disk scale
length, $h$. In this figure, the diagonal lines of constant luminosity 
intersect both HSB (filled squares) and LSB (open squares) points.  
At intermediate luminosity, there can be both HSB and LSB galaxies 
(a high central surface brightness and small $h$, or a low central surface brightness and large $h$).  The fact 
that two galaxies with the same luminosity can have wildly different 
surface brightnesses leads to the belief that some mechanism, perhaps 
related to the initial halo spin parameter, prevents LSB systems 
from collapsing to the same densities as HSBs (e.g. Dalcanton et al. 1997).

\Fig{csb_h} makes obvious that for a given luminosity, the LSB galaxy
must be more extended than the HSB galaxy.  With this small sample, 
no LSB galaxy
exceeds 10$^{11} L_{\odot}$. Applying the statistical F test to the
distribution of disk scale lengths yields a bimodality confidence level
of 60\%, implying that this distribution is not likely bimodal, as is
clear by eye.

\begin{figure*} 
\centering
\includegraphics[width=0.9\textwidth] {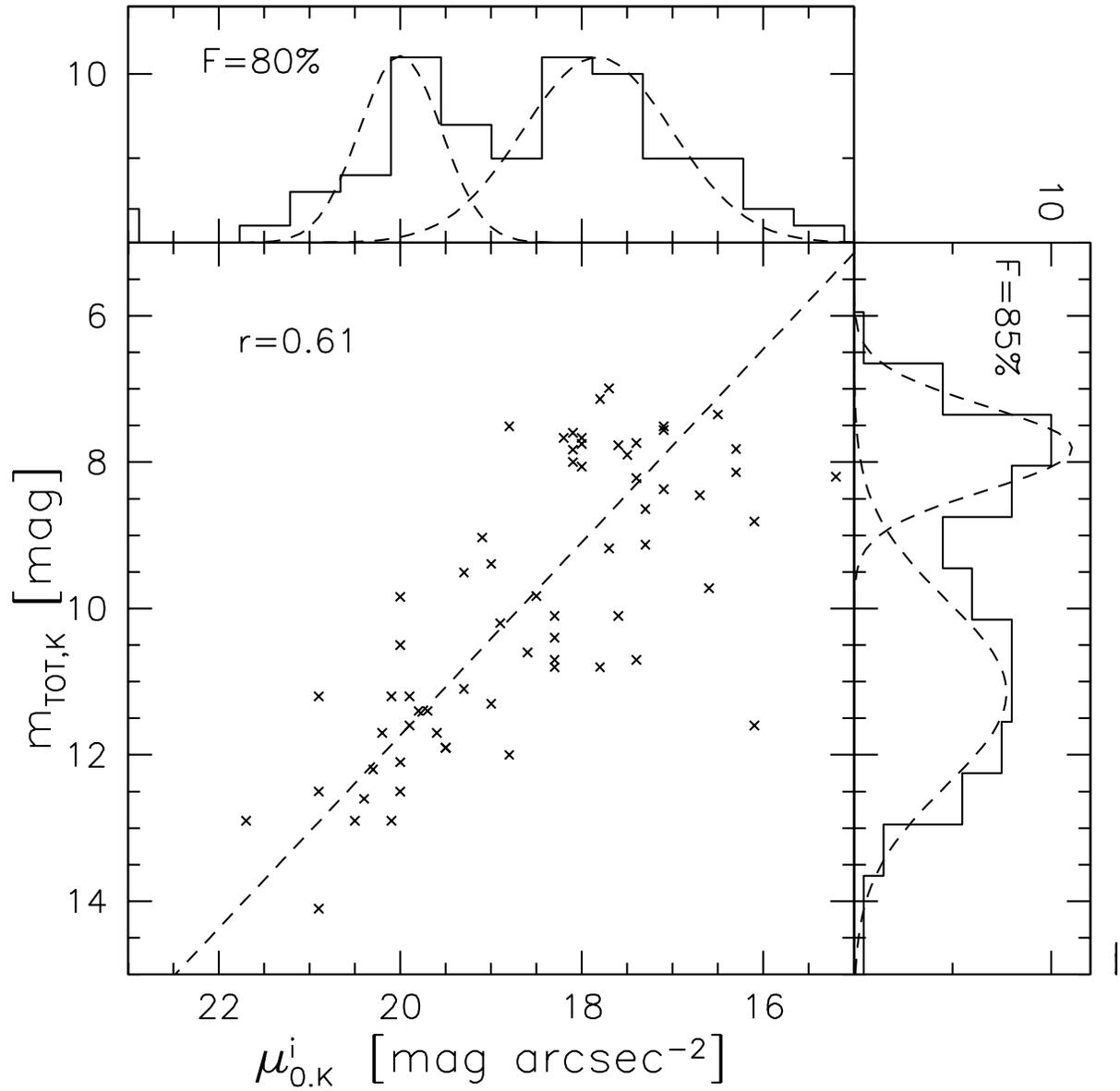}
\caption{Distribution of inclination-corrected $K^\prime$-band disk central 
surface brightnesses, $\mu_{0,K}^i$, as a function of total $K^\prime$-band
magnitude, $m^K_{TOT}$ for the sample of 65 UMa galaxies. It is
clear from the Pearson r coefficient of 0.61 that there is a 
correlation between $\mu_{0,K}^i$ and $m^K_{TOT}$.}
\label{fig:mu0-L}
\end{figure*}

We look in \Fig{mu0-L} for a possible correlation between the observed 
bimodality in $\mu_{0,K}c^i$ and the near-IR luminosity function. 
The distribution of total luminosities is strongly bimodal at $K^\prime$ band, 
even though 
the correlation between the disk central
surface brightness and the total luminosity of a galaxy is weak.
There appears to be a strong preference towards either $K^\prime$$_{TOT}$=8 mag 
or $K^\prime$$_{TOT}$=11 mag, with a large ($\sim$2 mag) gap with few 
galaxies. This bimodality in total magnitude is immediately 
apparent by eye, with an F-test yielding a confidence level of 85\%
\footnote{TV97 only discussed the luminosity function at B band 
which is flat rather than bimodal. High luminosity galaxies tend to be 
redder than low luminosity galaxies which increases the separation 
between these types, resulting in the observed minimum in the 
luminosity function at $K^\prime$ band.}. 
There also appears to be an almost equal mix of high and low surface 
brightness disks for bright galaxies, while faint galaxies are dominated 
by LSB disks.  However, it should be noted that several of the bright 
galaxies (primarily S0s) do not have pure exponential disks, 
and the measurement of $\muc$ in this case will yield a low and 
likely meaningless estimate.

\begin{figure*} 
\centering
\includegraphics[width=0.9\textwidth] {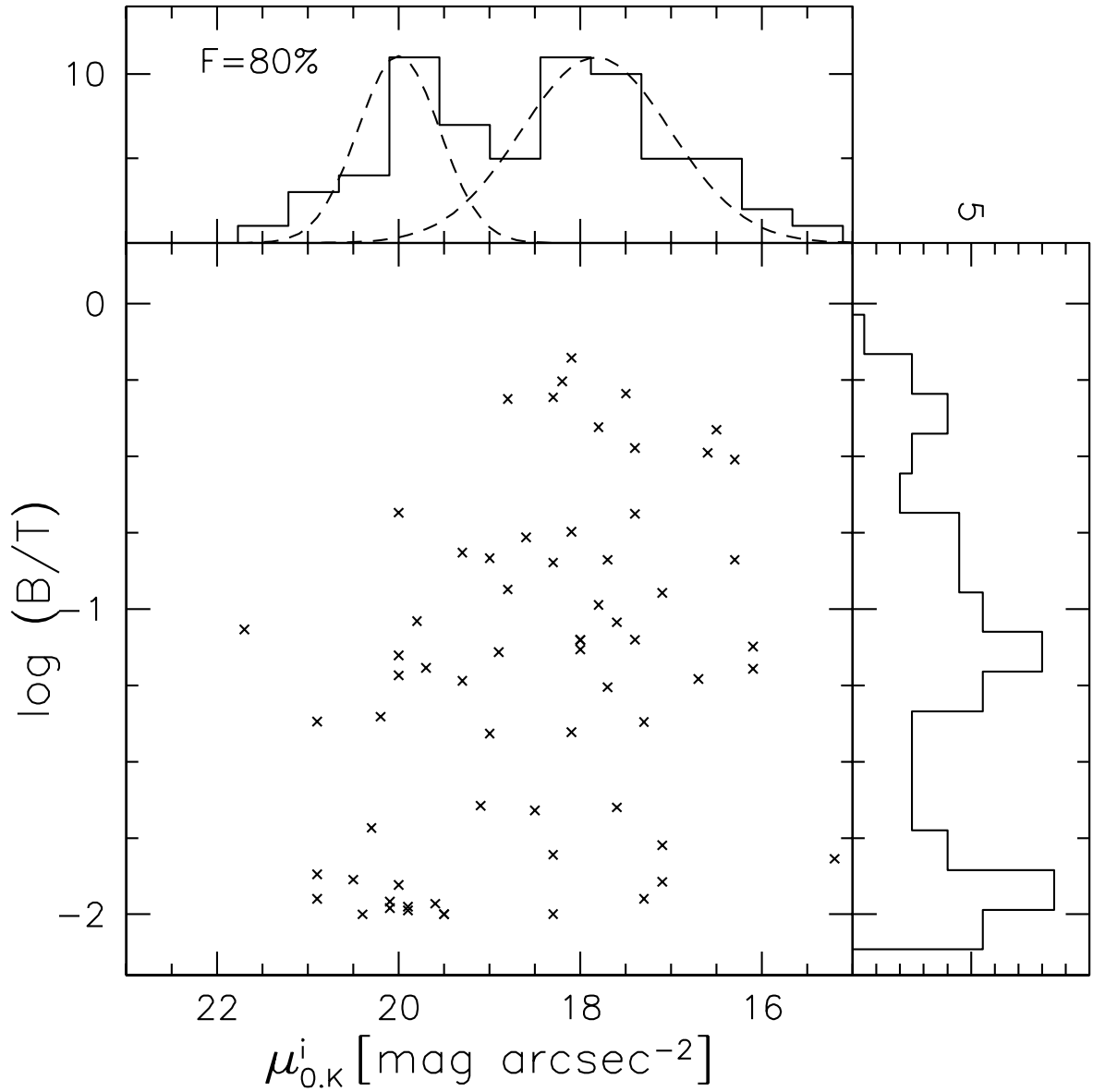}
\caption{Distribution of inclination-corrected $K^\prime$-band disk central 
surface brightnesses, $\mu_{0,K}^i$, as a function of bulge-to-total
ratio, $B/T$, for the sample of 65 UMa galaxies.}
\label{fig:mu0_btt}
\end{figure*}

It is natural to ask if the bimodality in the disk $\muc$ is 
correlated with any bulge structure.  \Fig{mu0_btt} shows
the disk central surface brightness as a function of the 
bulge-to-total ratio, $B/T$.  We can clearly assert that 
the surface brightness bimodality of disk galaxies and the 
well known observed color (or morphological) bimodality 
(e.g. Strateva et al. 2001) are independent.  Galaxies 
with large bulges (typically redder, e.g. S0-Sb) or galaxies
with little-to-no bulge (typically bluer, e.g. Sc-Irr) all 
exist in the form of HSB or LSB disks. 

\begin{figure*} 
\centering
\includegraphics[width=0.9\textwidth] {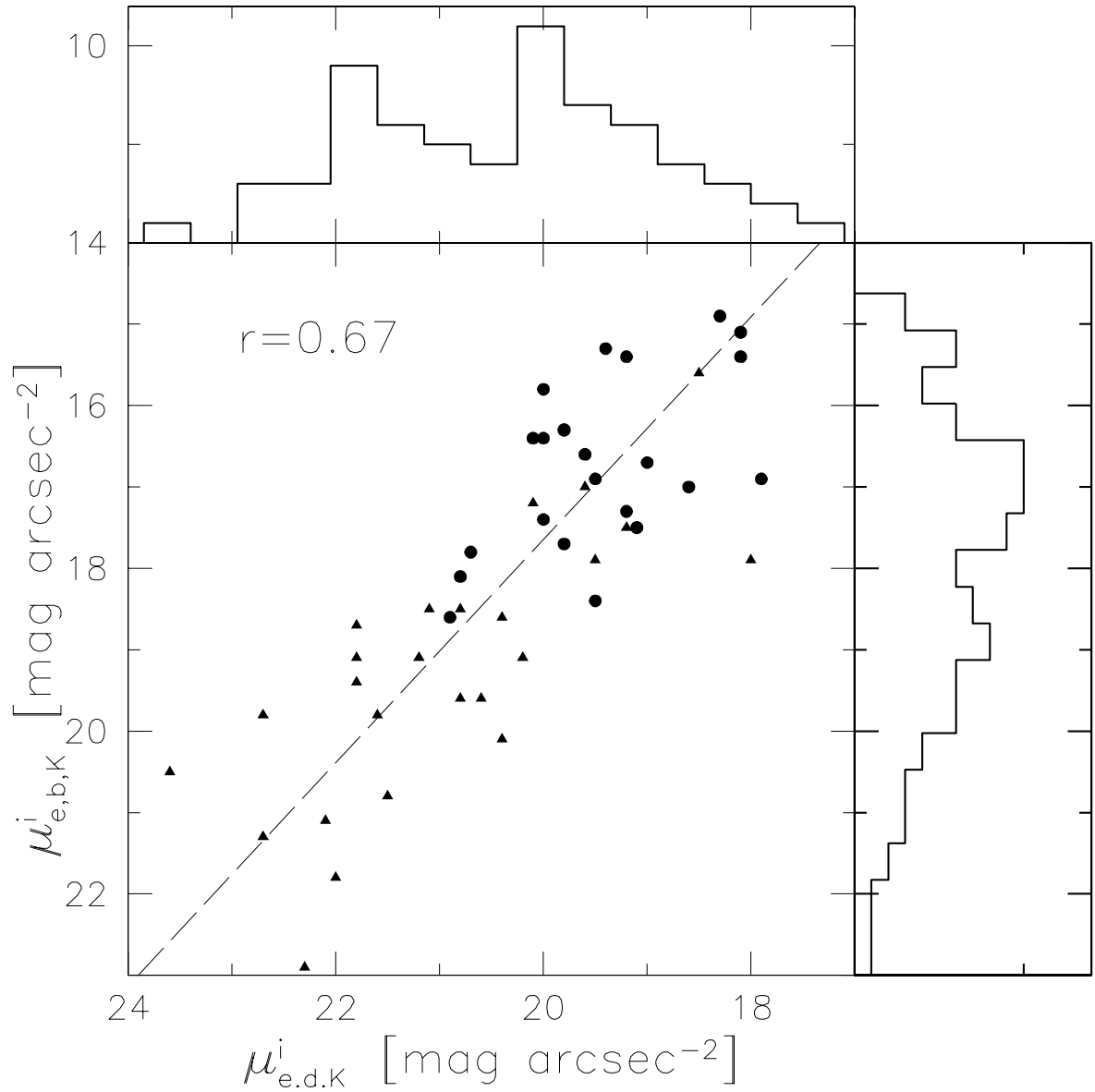}
\caption{Distribution of inclination-corrected $K^\prime$-band disk effective
surface brightnesses, $\mu_{e,d}^i$, as a function of bulge effective
surface brightness, $\mu_{e,b}$, for 48 UMa galaxies with discernible bulges. The circles refer to galaxies with
$K^\prime$$_{TOT}<$ 9.5, while the triangles refer to galaxies fainter than
this limit.}
\label{fig:mud_mub}
\end{figure*}

Another way to illustrate the connection between bulges 
and disks is shown in \Fig{mud_mub}. Here we
see a strong correlation between the bulge effective surface
brightness and the disk effective surface brightness (Pearson
r=0.67), using Eq.~\ref{eq:e-0} to convert from $\muc$ to $\mu_e$. 
The striking bimodality that is apparent in disk surface 
brightness is not seen in the bulge.
From this strong correlation, it is clear that bright disks tend to 
harbor bright bulges, while faint disks tend to harbor 
faint bulges (or none at all). 

\begin{figure*} 
\centering
\includegraphics[width=0.9\textwidth] {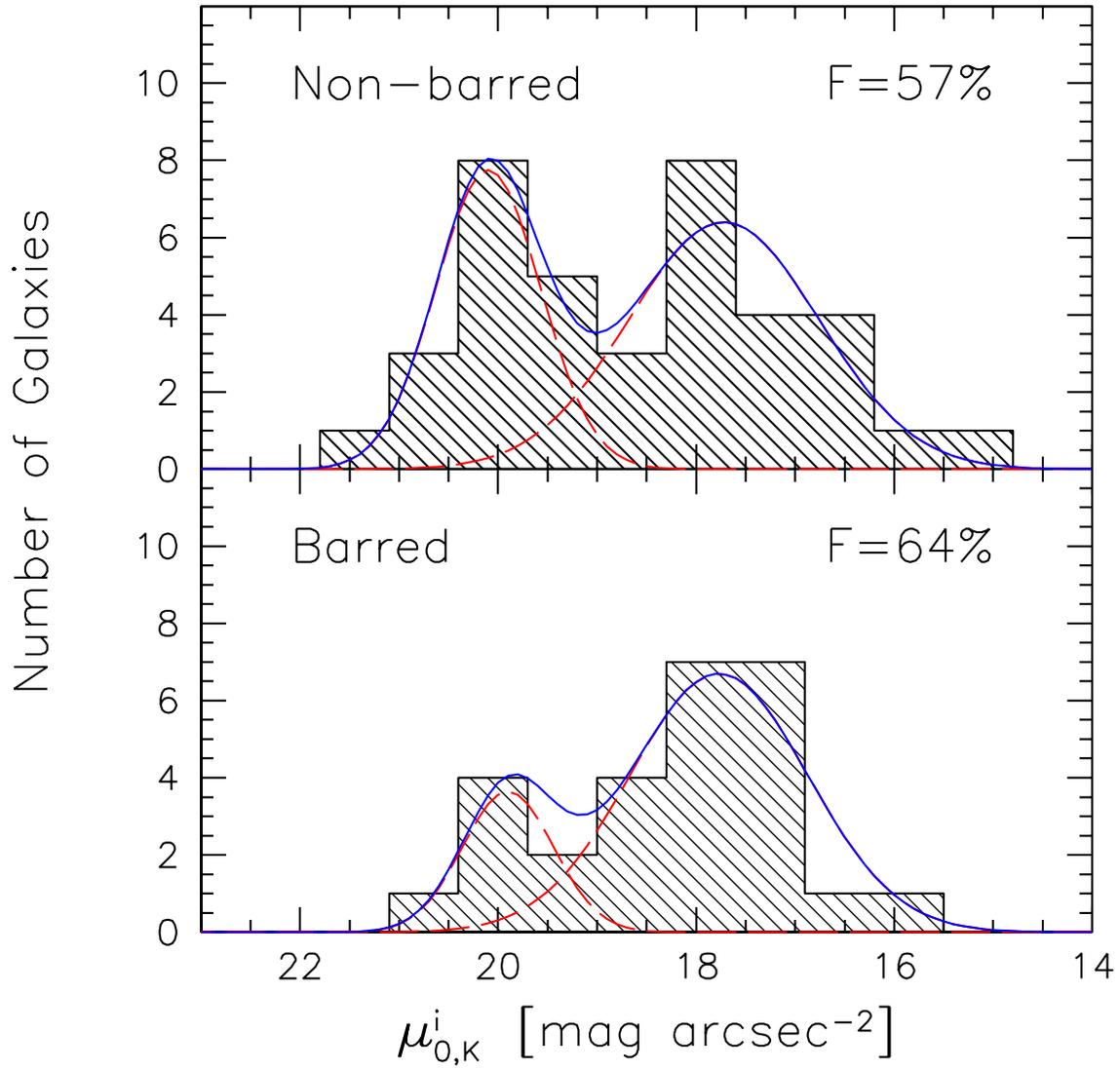}
\caption{Distribution of inclination-corrected $K^\prime$-band disk
  central surface brightnesses, $\mu_{0}^i$ for subsamples of barred
  and non-barred galaxies.}
\label{fig:csb_bar}
\end{figure*}

As mentioned in \se{data} we do not independently fit for galaxy bars in
our B/D decompositions. For weak bars, the bulge absorbs
most of their signal. We must however verify that stronger bars do not skew B/D parameters to the point of inducing a brightness bimodality. \Fig{csb_bar} shows the distribution of
$\muc$ for galaxies with and without a bar (the morphological type is taken from the Nasa Extragalactic Database [NED]). Bimodality is clearly seen
in both sub-samples, albeit with weaker F-statistics than for the full sample since the number of objects in each sub-samples is less. \Fig{csb_bar} shows that bimodality is not the result of bar structure.

\begin{figure*} 
\centering
\includegraphics[width=0.9\textwidth] {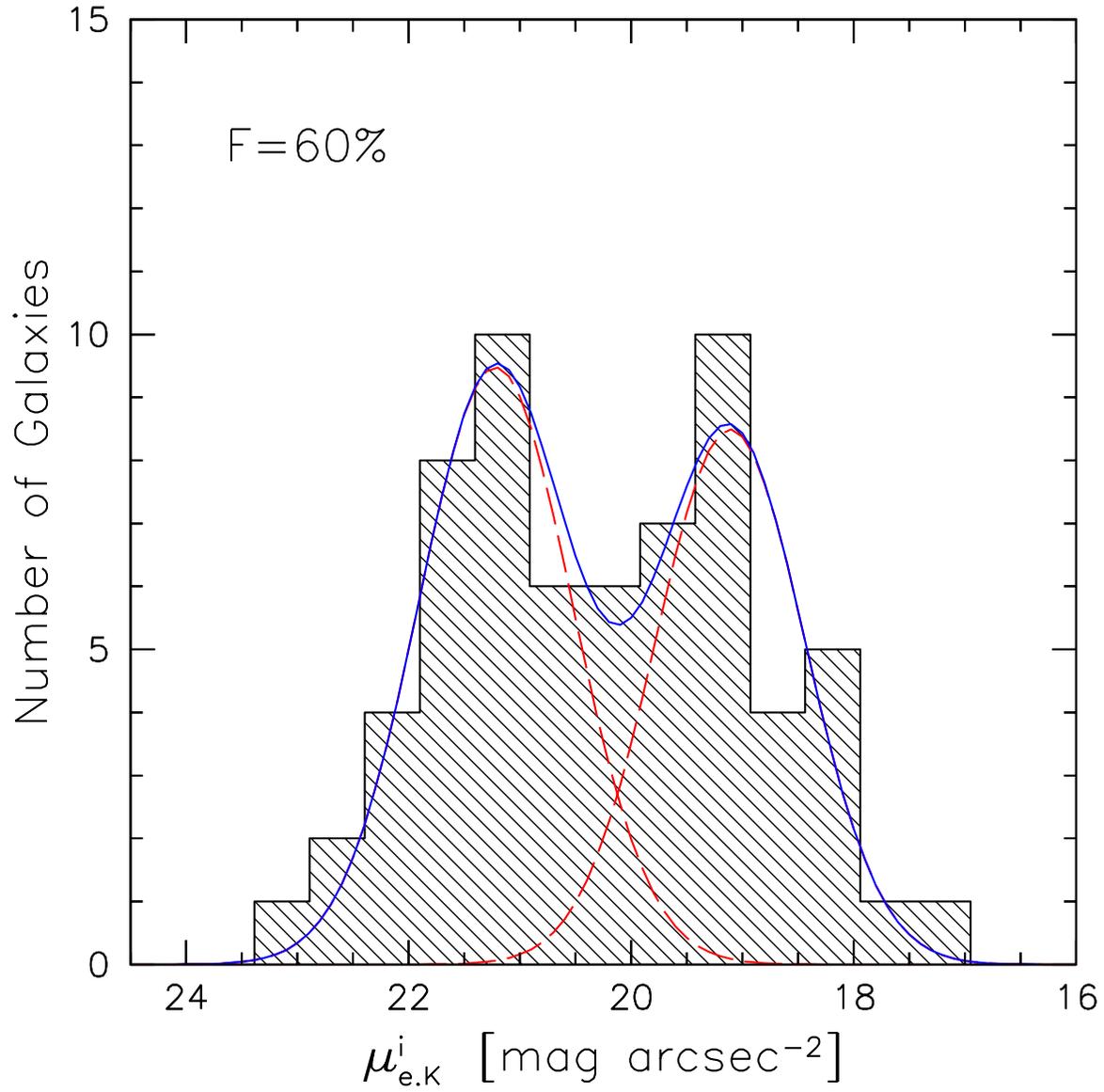}
\caption{Distribution of inclination-corrected $K^\prime$-band disk effective
surface brightnesses, $\mu_{e,d}^i$ as measured by fitting the disk
with a S\'ersic function for the sample of 65 UMa galaxies. The solid line represents the sum of the two
Gaussians (dashed lines).}
\label{fig:csb_sersic}
\end{figure*}

Finally, it is unneccessary to restrict ourselves to the 
choice of an exponential function to describe the disk 
of a galaxy.  We report in \Fig{csb_sersic} the results of
fitting two \Sersic functions to model the bulge and disk 
of the galaxy.  This more generalized approach preserves 
the bimodality of the disk surface brightness, giving a 
bimodality confidence level of 60\%.  
The peaks in this case are shifted by roughly 1.8 mag from those seen in 
\Fig{uma_csb}, which corresponds to the difference in 
measurement between central and effective surface brightness 
(Eq.~\ref{eq:e-0}). This fitting method also provides us with a way 
of testing the ubiquity of the exponential disk. We find 
that the ``disks'' in this sample have $<n>=0.8$
with $\sigma=0.3$. So, while the vast majority of our disks 
are nearly exponential (see \Fig{ndist}), there 
is a broad distribution in shapes that are best described with
B/D profile decompositions. 

\section{A Dynamical Explanation?}
The two surface brightness peaks identify two distinct regimes
of stellar surface density.  Could this structural bimodality 
be linked to any dynamical process?  It is known that HSB and 
LSB galaxies have different rotation curve characteristics. 
HSB galaxies typically achieve maximal rotation speed at or
within $r=2.15h$ (e.g. Courteau 1997) while the rotation curves
of LSB galaxies are still rising at that radius. For HSB galaxies
the baryonic disk component may substantially dominate the halo
interior to $2h$ while for LSBs, the baryons are sub-dominant 
to the dark matter component at nearly all galactocentric radii.
This conclusion is supported by mass modeling and dynamical 
arguments (e.g. Courteau \& Rix 1999: Dutton \etal 2005; Dutton 
\etal 2007). The proposed explanation offered by TV97 for this 
phenomenon was that of two stable radial configurations for the 
baryons in galaxies, an idea proposed by Mestel (1963). The
configuration occupied by the LSB galaxies would arise from 
small-amplitude ($\sim 1\sigma$) initial fluctuations and thus 
that have formed late with high angular momentum.  Material 
from the disk in these systems is not sufficiently funneled 
to the center due to their high angular momentum to ever 
dominate the dark halo (Dalcanton \etal 1997).  The alternate 
configuration involves galaxies either with low initial angular 
momentum or with baryon content that loses angular momentum 
through interactions.  Material with minimal angular momentum 
can collapse far enough to dominate the dark halo in the inner 
regions.  There would be a threshold amount of angular momentum 
that determines whether the disk collapses to a HSB configuration 
or is unable to fully collapse and ``stalls'' as an LSB galaxy.

\begin{figure*}
\centering
\includegraphics[width=0.9\textwidth] {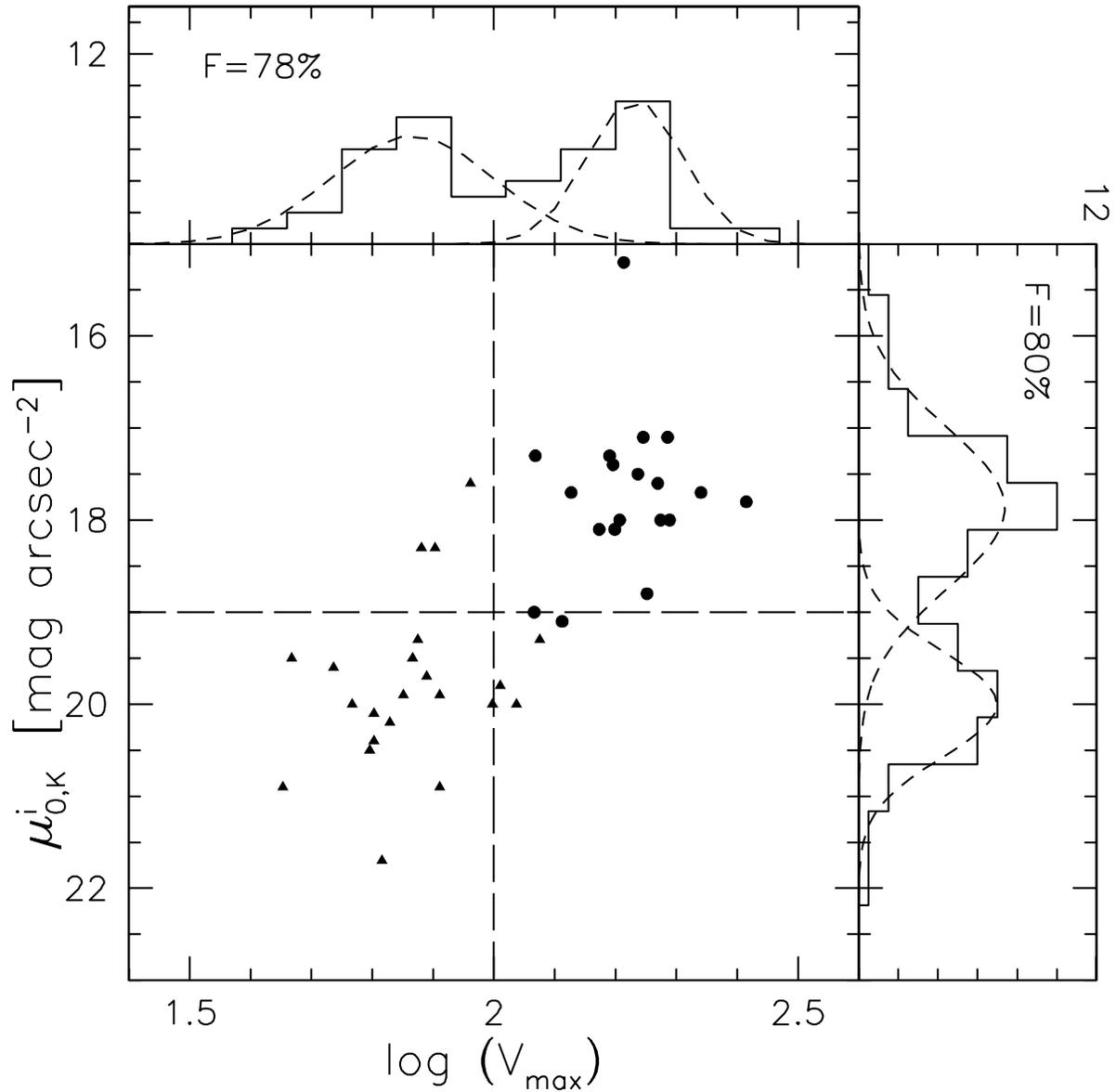}
\caption{Distribution of inclination-corrected $K^\prime$-band disk central 
surface brightnesses, $\mu_{0,K}^i$, as a function of the maximum
rotational velocity, $V_{max}$, for 62 UMa galaxies with rotational velocities. The circles refer to galaxies with
$K^\prime$$_{TOT}<$ 9.5 mag, while the triangles refer to galaxies fainter
than this cutoff.}
\label{fig:mu0-v}
\end{figure*}

In order to assess any dynamical connection with surface 
brightnesses, we obtained \ion{H}{I} line widths from Marc 
Verheijen (Verheijen \& Sancisi 2001) for 62 UMa galaxies.  
The distribution of rotational velocities, corrected for 
inclination, for those galaxies is shown in \Fig{mu0-v}. 
Not only is there a clear bimodality in the $\muc$ distribution,
but there is also one in the distribution of rotational velocities.
Yet the correlation between rotational velocity and surface 
brightness is relatively weak ($r=0.7$).  By virtue of the 
Tully-Fisher relation (Tully \& Fisher 1977), this is expected 
from \Fig{mu0-L}.  There is a minor concern that LSB galaxies 
may reach a maximum rotation velocity at larger physical radii 
than HSB galaxies of the same luminosity (TV97).  A contribution 
toward bimodality in $V_{max}$ could thus arise from not sampling
the rotation curve far enough for LSB galaxies.  If our $V_{max}$ 
term is calculated at a limiting isophotal radius (say, $r_{max}$),
then it is plausible that the HSB galaxies will have $V(r_{max})=V_{max}$ 
whereas the LSB galaxies only have $V(r_{max})=aV_{max}$ where 
potentially $a<1$ in some cases.

We explore in greater detail in our companion paper on the light profiles 
and luminosity function of Virgo cluster galaxies (McDonald \etal 2008) 
the suite of possible scenarios that may explain the observed structural 
bimodality in cluster galaxies, including cluster-induced morphological 
transitions, gas depletion and/or dynamical thresholds in a cluster 
environment. 

Unlike traditional rich clusters such as Virgo, with a space density 
of $\sim$100 Mpc$^{-3}$, UMa closely resembles the field with a space 
density of only 8 Mpc$^{-3}$ (TV97; Trentham \& Tully 2002). 
Furthermore, we have confirmed TV97's finding that the surface 
brightness bimodality is enhanced when only the most isolated 
galaxies in UMa are considered.  This suggests that the intermediate 
surface brightness gap may be filled by galaxies found preferentially 
in dense environments: dwarf and giant ellipticals.  A follow-up study 
in a rich cluster will allow a complete morphological sampling, 
ensuring that bimodality is not due to the absence of some classes 
of galaxy.

Follow-up near-IR surveys of both field and cluster populations 
will be needed to determine the environmental dependence on the 
structural bimodality.  Previous studies of the surface brightness
distribution in the field have relied on optical data which requires
a complicated, and not fully understood, deprojection of the surface
brightness profiles to the face-on case.  The near-IR field galaxy 
surveys collected to date have traditionally been too shallow 
to properly sample the LSB peak (e.g. de Jong 1996; Jarrett \etal 
2003; Grauer \etal 2003; M03).  A deep, near-IR survey 
of a larger cluster such as Virgo, Fornax or Coma is needed to 
assess the cluster dependence of our result and whether UMa's surface 
brightness distribution is typical.   Since galaxy clusters are 
confined structures, they are an ideal laboratory for a study
of this type since volume corrections are less significant.  Note
that none of the nearby clusters listed above will be covered in 
any deep fashion by the ongoing UKIDSS survey\footnote{\tt http://www.ukidss.org/}. 

Finally, though more expensive observationally than a simple photometric 
investigation, a complete dynamical study of a cluster environment will 
ultimately be needed to pin down the cause of structural bimodality.  
Galaxy line widths are currently available mostly for HSB galaxies and
their sources are often heterogeneous.  A dedicated dynamical survey down
to low rotational velocities could not only assess the bimodality of 
the velocity function of cluster galaxies, in conjunction with that 
of the luminosity function, but also examine whether those functions 
are related in any way to the shape of the rotation curve and/or the 
distribution of visible to dark matter in galaxies. 

\section{Conclusion}
We have implemented a new code for bulge-disk decompositions 
which accounts for compact nuclei, seeing, spiral arms and disk 
truncations. We also allow for both exponential and S\'ersic 
bulges and disks. By carefully fitting all the components of 
a given galaxy, we can extract confident estimates of the 
various galaxy structural parameters.

We have confirmed the bimodality of disk central surface brightness
as reported by TV97 and, in addition, find a distinct minimum in the 
near-IR luminosity function at $K^{\prime} \simeq 9$ ($M_K \sim -22$)
and a corresponding minimum in the distribution of maximal rotational 
velocities for UMa galaxies.  The concern about small number statistics
is, however, real (Bell \& de Blok 2000).  Furthermore, the UMa cluster
population resembles that of the field and a sample that includes more 
early type galaxies would offset any ambiguities arising from morphological
dependencies. A new near-IR based study of a richer, denser, more evolved
cluster is needed (see McDonald \etal 2008).  Likewise, a 
statistically complete sample of field galaxies would address possible
environmental dependence of the observed galaxy structural bimodality. 

\section{Acknowledgements}
We acknowledge Lauren MacArthur for useful conversations and suggestions 
about bulge-disk decompositions and Marc Verheijen for providing the 
raw Ursa Major photometry and line widths. We thank the referee,
Enrico Maria Corsini, for his careful read of the paper and the suggestion to confirm that bimodality is not linked to galaxy bars. SC would like to acknowledge 
financial support via a Discovery Grant from the National Science and 
Engineering Council of Canada. RBT acknowledges support from US National 
Science Foundation award AST 03-07706.

This research has made use of $(i)$ the NASA/IPAC Extragalactic Database 
(NED) which is operated by the Jet Propulsion Laboratory, California 
Institute of Technology, under contract with the National Aeronautics 
and Space Administration, as well as NASA's Astrophysics Data System; 
$(ii)$ the $Sloan$ $Digital$ $Sky$ $Survey$ (SDSS). Funding for the 
creation and distribution of the SDSS Archive has been provided by 
the Alfred P. Sloan Foundation, the Participating Institutions, the 
National Aeronautics and Space Administration, the National Science 
Foundation, the U.S. Department of Energy, the Japanese Monbukagakusho, 
and the Max Planck Society. The SDSS Web site is http://www.sdss.org/. 
The SDSS is managed by the Astrophysical Research Consortium (ARC) for 
the Participating Institutions. 
$(iii)$ the HyperLeda database (http://leda.univ-lyon1.fr).


\clearpage

\onecolumn

\begin{longtable}{rcrrrrrrrrrrrr}
\caption[]{B/D Parameters}\\
\hline
\hline
\multicolumn{1}{c}{(1)}&  \multicolumn{1}{c}{(2)}&
\multicolumn{1}{c}{(3)}&     \multicolumn{1}{c}{(4)}&
\multicolumn{1}{c}{(5)}&     \multicolumn{1}{c}{(6)}&
\multicolumn{1}{c}{(7)}&     \multicolumn{1}{c}{(8)}&
\multicolumn{1}{c}{(9)}&     \multicolumn{1}{c}{(10)}&
\multicolumn{1}{c}{(11)}&    \multicolumn{1}{c}{(12)}&
\multicolumn{1}{c}{(13)}&    \multicolumn{1}{c}{(14)}\\
\multicolumn{1}{c}{PGC}&   
\multicolumn{1}{c}{$T$}&
\multicolumn{1}{c}{$K'_T$}&
\multicolumn{1}{c}{$b/a$}&
\multicolumn{1}{c}{$B/T$}&
\multicolumn{1}{c}{$C_{28}$}&
\multicolumn{1}{c}{$\mu_e^{K'}$}&
\multicolumn{1}{c}{$r_e^{K'}$}&
\multicolumn{1}{c}{$\mu_{e,d}^{K'}$}&
\multicolumn{1}{c}{$r_{e,d}^{K'}$}&
\multicolumn{1}{c}{$\mu_{e,b}^{K'}$}&
\multicolumn{1}{c}{$r_{e,b}^{K'}$}&
\multicolumn{1}{c}{$n_{b}^{K'}$}&
\multicolumn{1}{c}{$\mu_{nuc}^{K'}$}\\
\hline
\\
\endhead
34971&   Sm&   11.22&   0.31&   0.00&   2.97&   21.47&   31.81&   20.47&   38.32&   &&&   17.98\\
35202&   Sd&   11.36&   0.98&   0.05&   2.30&   20.94&   19.20&   21.53&   30.99&   20.77&   8.56&   0.30&   17.42\\
35676&   SBc&   7.83&   0.58&   0.03&   1.93&   19.81&   73.65&   19.37&   76.97&   17.44&   5.87&   0.60&   14.77\\
35711&   SBa&   8.64&   0.56&   0.03&   2.78&   18.79&   32.22&   18.48&   35.51&   17.52&   3.75&   0.40&   14.31\\
35999&   SBb&   9.13&   0.33&   0.00&   2.91&   18.83&   24.62&   17.94&   29.06&   &&&   15.44\\
36136&   SBcd&   10.77&   0.37&   0.00&   2.44&   20.01&   20.56&   19.05&   23.32&   &&&   21.92\\
36343&   Scd&   11.17&   0.11&   0.03&   3.12&   21.88&   40.70&   20.34&   65.16&   19.80&   10.91&   0.40&   17.68\\
36528&   Sm&   11.74&   0.74&   0.03&   2.42&   21.15&   22.71&   21.64&   32.92&   21.80&   9.42&   0.50&   18.60\\
36686&   S0&   10.79&   0.58&   0.38&   3.52&   18.36&   8.21&   19.00&   13.18&   16.97&   5.19&   0.50&   14.05\\
36699&   Sc&   7.75&   0.20&   0.07&   4.16&   18.79&   47.25&   18.09&   64.43&   16.28&   3.88&   0.70&   14.61\\
36825&   Im&   11.73&   0.53&   0.00&   2.60&   20.89&   21.07&   20.72&   25.63&   &&&   18.08\\
36875&   SBc&   7.51&   0.75&   0.48&   4.09&   19.09&   46.54&   20.35&   82.48&   17.80&   23.34&   1.20&   13.68\\
36897&   SB0/a&   11.31&   0.63&   0.14&   3.50&   20.00&   13.49&   20.34&   21.61&   18.51&   4.58&   0.40&   17.17\\
36953&   SBd&   10.43&   0.93&   0.13&   1.96&   20.37&   24.25&   20.09&   25.43&   19.06&   12.50&   0.20&   16.84\\
36990&   S0&   11.64&   0.78&   0.07&   3.08&   17.58&   4.20&   17.69&   5.12&   17.92&   2.05&   0.10&   16.76\\
37024&   Sd&   10.59&   0.85&   0.16&   3.74&   19.73&   17.25&   20.25&   23.97&   18.62&   4.76&   1.00&   17.28\\
37036&   Scd&   9.03&   0.24&   0.01&   2.68&   20.26&   54.91&   19.33&   64.37&   18.64&   4.02&   0.20&   17.20\\
37037&   IBm&   11.64&   0.93&   0.00&   1.73&   22.00&   23.23&   21.61&   26.20&   &&&   19.29\\
37038&   SB?&   11.86&   0.46&   0.00&   2.76&   20.99&   19.48&   20.44&   22.91&   &&&   22.96\\
37073&   S0&   10.65&   0.82&   0.19&   4.58&   18.64&   9.48&   18.96&   13.29&   17.51&   2.46&   0.60&   12.23\\
37136&   Sb&   9.72&   0.97&   0.31&   4.65&   17.54&   7.52&   18.42&   13.30&   15.63&   2.49&   0.70&   12.46\\
37164&   SBm&   12.85&   0.29&   0.00&   2.31&   21.59&   17.19&   20.60&   21.03&   &&&   19.66\\
37217&   Sm&   12.08&   0.81&   0.00&   2.31&   21.27&   21.44&   21.54&   27.16&   &&&   17.39\\
37229&   Sc&   7.74&   0.88&   0.07&   3.24&   18.96&   46.66&   19.07&   55.55&   17.30&   6.48&   0.80&   15.44\\
37290&   Sbc&   8.45&   0.63&   0.05&   3.31&   18.42&   27.55&   18.05&   29.81&   16.95&   4.77&   0.30&   15.27\\
37306&   SBbc&   6.99&   0.45&   0.13&   3.70&   18.79&   60.11&   18.61&   83.48&   16.87&   11.31&   1.50&   14.09\\
37418&   Scd&   12.58&   0.16&   0.00&   2.45&   21.84&   21.85&   20.20&   26.09&   &&&   26.40\\
37466&   Sbc&   9.39&   0.24&   0.03&   3.13&   20.00&   39.22&   19.22&   54.36&   18.12&   4.35&   0.70&   16.22\\
37520&   SBb&   8.81&   0.86&   0.05&   2.36&   17.64&   17.03&   17.76&   18.28&   16.92&   2.61&   0.20&   11.83\\
37525&   SBm&   10.50&   0.52&   0.05&   2.55&   21.14&   38.80&   21.11&   52.49&   19.38&   10.18&   0.20&   17.92\\
37542&   SBm&   10.07&   0.72&   0.08&   3.31&   19.44&   19.12&   19.10&   21.00&   17.87&   4.33&   0.20&   16.79\\
37550&   Scd&   12.00&   0.75&   0.11&   3.11&   20.02&   10.61&   20.26&   14.20&   19.60&   4.66&   0.30&   17.93\\
37553&   Im&   11.14&   0.47&   0.14&   3.74&   19.82&   14.59&   20.36&   26.41&   19.08&   5.62&   0.60&   17.85\\
37584&   SBd&   10.18&   1.00&   0.06&   2.40&   20.62&   30.17&   20.75&   38.88&   19.57&   8.97&   0.30&   17.95\\
37617&   SBbc&   7.14&   0.54&   0.09&   3.68&   19.70&   78.95&   18.97&   92.48&   16.61&   8.42&   1.10&   13.07\\
37621&   Scd&   14.08&   0.24&   0.00&   3.62&   21.93&   10.82&   21.22&   16.25&   &&&   19.98\\
37642&   S0&   7.35&   0.82&   0.38&   5.33&   16.87&   15.68&   18.09&   34.77&   14.87&   6.12&   0.80&   11.38\\
37682&   SBm&   12.52&   0.57&   0.00&   2.16&   22.15&   22.53&   22.10&   31.93&   21.32&   1.67&   0.20&   18.74\\
37691&   Sb&   7.77&   0.16&   0.01&   3.20&   18.70&   45.93&   17.38&   53.81&   &&&   12.66\\
37692&   SBcd&   9.83&   0.72&   0.01&   2.48&   19.98&   31.15&   20.00&   39.11&   20.06&   5.22&   0.40&   17.10\\
37697&   SBd&   9.51&   0.12&   0.05&   2.91&   20.16&   38.27&   18.82&   63.82&   18.48&   16.78&   0.20&   16.88\\
37700&   Im&   12.92&   0.38&   0.00&   2.14&   21.49&   16.36&   21.27&   24.04&   22.87&   1.87&   0.10&   18.31\\
37719&   Sab&   8.20&   0.40&   0.01&   4.15&   16.81&   15.07&   16.05&   16.93&   &&&   13.70\\
37735&   SBcd&   11.36&   0.45&   0.08&   2.68&   21.28&   25.92&   20.74&   32.01&   19.82&   9.53&   0.40&   17.90\\
37760&   S0&   7.67&   0.22&   0.55&   5.91&   16.86&   16.25&   18.40&   53.66&   16.40&   13.31&   1.80&   13.28\\
38068&   SBbc&   8.00&   0.51&   0.17&   4.73&   19.92&   63.56&   19.23&   63.66&   16.35&   4.99&   1.70&   12.93\\
38283&   SBc&   9.18&   0.23&   0.05&   3.87&   18.66&   21.76&   17.96&   29.61&   18.43&   4.96&   0.90&   16.02\\
38302&   SBbc&   7.56&   0.43&   0.01&   2.82&   19.30&   63.63&   18.00&   49.63&   &&&   12.93\\
38356&   Sdm&   11.22&   0.25&   0.00&   2.33&   21.36&   33.01&   20.45&   45.96&   &&&   18.43\\
38370&   SBb&   8.06&   0.31&   0.07&   2.94&   19.46&   57.87&   18.49&   55.11&   16.34&   4.47&   0.40&   14.36\\
38375&   Sdm&   11.91&   0.38&   0.00&   2.62&   20.93&   19.31&   20.23&   21.96&   &&&   25.25\\
38392&   SBb&   7.90&   0.57&   0.50&   6.24&   16.95&   12.83&   18.76&   37.34&   15.30&   6.52&   0.80&   11.25\\
38440&   S0&   7.60&   0.22&   0.66&   5.69&   16.08&   11.13&   18.35&   47.74&   15.75&   12.46&   1.70&   12.37\\
38503&   S0&   10.05&   0.43&   0.48&   4.32&   18.14&   10.62&   19.22&   21.26&   17.17&   7.90&   0.80&   12.35\\
38507&   S0&   12.45&   0.60&   0.20&   3.52&   20.78&   10.59&   21.25&   19.89&   19.09&   4.78&   0.60&   17.83\\
38643&   S0&   8.14&   0.62&   0.14&   4.16&   17.75&   22.00&   17.60&   26.39&   15.37&   3.19&   0.70&   11.37\\
38654&   SB0&   7.82&   0.54&   0.30&   5.18&   16.82&   13.99&   17.44&   26.92&   15.06&   4.47&   1.10&   13.52\\
38795&   SBb&   7.51&   0.17&   0.00&   3.54&   18.96&   55.18&   16.96&   52.03&   &&&   13.84\\
38951&   Im&   12.18&   0.61&   0.01&   1.30&   21.85&   25.94&   21.59&   31.27&   21.09&   3.85&   0.10&   17.52\\
38988&   Scd&   9.84&   0.14&   0.06&   3.52&   20.55&   38.91&   19.70&   67.94&   18.68&   8.85&   0.70&   16.71\\
39237&   Sa&   10.71&   0.58&   0.01&   2.40&   19.21&   14.27&   19.48&   18.39&   &&&   15.63\\
39241&   Sb&   7.67&   0.17&   0.06&   3.23&   19.62&   67.27&   17.90&   71.35&   17.72&   14.89&   0.90&   14.56\\
39285&   S0&   8.37&   0.29&   0.10&   4.16&   18.06&   24.81&   17.63&   33.14&   16.67&   4.39&   1.00&   14.81\\
39344&   Sdm&   12.90&   0.15&   0.08&   3.20&   22.43&   24.08&
21.52&   37.93&   20.50&   7.21&   0.30&   18.43\\
40228&   SB0&   8.22&   0.32&   0.33&   5.75&   17.17&   13.56& 17.96&
35.92&   15.37&   4.90&   0.80&   12.78\\

\end{longtable}

Table captions -- Col.(1) PGC number; Col.(2) Morphology from the NASA
Extragalactic Database; Col.(3) Total $K^\prime$ magnitude;
Col.(4) Ratio of the minor-to-major axis diameters; Col.(5) Ratio of
the bulge to the total luminosity; Col.(6) Galaxy concentration,
defined as $C_{28}=5\log(r_{80}/r_{20})$ where $r_{20}$ and $r_{80}$
are radii enclosing 20\% and 80\% of the total light, respectively;
Col.(7) Effective surface brightness, defined as the surface
brightness at the galaxy half-light radius; Col.(8) Effective radius
within which half of the total galaxy light is enclosed; Col.(9)
Effective surface brightness of the disk; Col.(10) Half-light radius
of the disk; Col.(11) Effective surface brightness of the bulge;
Col.(12) Half-light radius of the bulge; Col.(13) \Sersic parameter
for the bulge; Col.(14) Surface brightness of the galaxy nucleus.

\clearpage

\newpage



\clearpage 


\clearpage 


\clearpage 


\clearpage 


\clearpage 

\clearpage

\clearpage

\clearpage 


\clearpage 

\clearpage 


\clearpage 


\clearpage 


\clearpage 


\clearpage 


\clearpage 


\clearpage 


\end{document}